%% file: preprint.tex
\documentclass[10pt,a4paper,twocolumn]{article}

\usepackage[a4paper,top=2.1cm,bottom=2.3cm,left=1.75cm,right=1.75cm,columnsep=0.65cm,headsep=0.5cm,footskip=0.9cm]{geometry}
\usepackage[T1]{fontenc}
\usepackage[utf8]{inputenc}
\usepackage{newtxtext,newtxmath}
\usepackage[scaled=0.86]{beramono}
\usepackage{microtype}
\DisableLigatures[i]{encoding = *}
\usepackage{amsmath}
\usepackage{booktabs,tabularx,array,multirow}
\usepackage{threeparttable}
\usepackage{enumitem}
\usepackage[dvipsnames]{xcolor}
\usepackage{caption}
\usepackage{listings}
\usepackage{etoolbox}
\usepackage{placeins}
\usepackage{dblfloatfix}
\usepackage{tikz}
\usetikzlibrary{arrows.meta,positioning,calc}
\usepackage{pgfplots}
\pgfplotsset{compat=1.18}
\usepackage[authoryear,round]{natbib}
\usepackage{titlesec}
\usepackage{fancyhdr}
\usepackage{xurl}
\definecolor{linkblue}{RGB}{0,72,140}
\usepackage[colorlinks,linkcolor=linkblue,citecolor=linkblue,urlcolor=linkblue]{hyperref}
\usepackage[capitalise,noabbrev,nameinlink]{cleveref}
\usepackage{orcidlink}

\input{preamble}

\titleformat{\section}{\normalfont\large\bfseries}{\thesection}{0.6em}{}
\titleformat{\subsection}{\normalfont\normalsize\bfseries}{\thesubsection}{0.6em}{}
\titleformat{\subsubsection}{\normalfont\normalsize\bfseries\itshape}{\thesubsubsection}{0.6em}{}
\titleformat{\paragraph}[runin]{\normalfont\normalsize\bfseries}{}{0pt}{}[.]
\titlespacing*{\section}{0pt}{2.0ex plus .6ex minus .3ex}{1.0ex plus .2ex}
\titlespacing*{\subsection}{0pt}{1.6ex plus .5ex minus .3ex}{0.7ex plus .2ex}
\titlespacing*{\subsubsection}{0pt}{1.3ex plus .4ex minus .2ex}{0.5ex}
\titlespacing*{\paragraph}{0pt}{0.9ex plus .3ex minus .2ex}{0.6em}

\renewenvironment{table}[1][]{\begin{table*}[!t]}{\end{table*}}
\renewenvironment{figure}[1][]{\begin{figure*}[!t]}{\end{figure*}}

\newcommand{\shorttitle}{Tokenised Sukuk: Asset Evidence and Programmable Shariah Compliance}
\fancypagestyle{firstpage}{%
  \fancyhf{}%
  \fancyhead[L]{\footnotesize\itshape Preprint. This version has not been peer reviewed.}%
  \fancyhead[R]{\footnotesize\itshape October 2026}%
  \fancyfoot[C]{\footnotesize\thepage}%
  }

\hypersetup{
  pdftitle={Tokenised Sukuk: A Verifiable Architecture for Asset Evidence and Programmable Shariah Compliance},
  pdfauthor={Rischan Mafrur, Khadijah, Gun Gun Febrianza},
  pdfkeywords={tokenized sukuk, asset backing, Shariah compliance, smart contract, verifiable attestation, digital finance}}

\begin{document}

\twocolumn[{%
\thispagestyle{firstpage}%
\begin{center}
{\LARGE\bfseries Tokenised Sukuk: A Verifiable Architecture for Asset\\[2pt] Evidence and Programmable Shariah Compliance\par}
\vspace{12pt}
{\large
Rischan Mafrur\,\orcidlink{0000-0003-4424-3736}\textsuperscript{1,$\star$}\qquad
Khadijah\textsuperscript{2}\qquad
Gun Gun Febrianza\textsuperscript{3}\par}
\vspace{8pt}
{\small
\textsuperscript{1}School of Business and Economics, Universiti Brunei Darussalam, Brunei Darussalam\\
\textsuperscript{2}CryptocoinHalal.com, Indonesia\\
\textsuperscript{3}Adaptive Institute, Indonesia\\[3pt]
\textsuperscript{$\star$}Corresponding author: \href{mailto:rischan.mafrur@ubd.edu.bn}{rischan.mafrur@ubd.edu.bn}\par}
\end{center}
\vspace{6pt}
\noindent\rule{\textwidth}{0.6pt}\par\vspace{4pt}
\noindent{\bfseries Abstract}\par\vspace{2pt}
\noindent\input{sections/abstract_preprint}\par
\vspace{5pt}
\par
\vspace{4pt}
\noindent\rule{\textwidth}{0.6pt}\par
\vspace{14pt}
}]

\input{sections/introduction}
\input{sections/background}
\input{sections/design}
\input{sections/evaluation}
\input{sections/discussion}
\input{sections/conclusion}
\FloatBarrier
\input{sections/declarations}

\bibliographystyle{elsarticle-harv}
\bibliography{references}

\onecolumn
\renewenvironment{table}[1][]{\begin{origtable}[!htbp]}{\end{origtable}}
\renewenvironment{figure}[1][]{\begin{origfigure}[!htbp]}{\end{origfigure}}
\appendix
\counterwithin*{table}{section}
\renewcommand{\thetable}{\Alph{section}.\arabic{table}}
\renewcommand{\theHtable}{\Alph{section}.\arabic{table}}
\input{sections/abbreviations}

\clearpage
\counterwithout*{table}{section}
\setcounter{section}{0}\setcounter{table}{0}\setcounter{figure}{0}\setcounter{lstlisting}{0}
\renewcommand{\thesection}{S\arabic{section}}
\renewcommand{\theHsection}{S\arabic{section}}
\renewcommand{\thetable}{S\arabic{table}}
\renewcommand{\theHtable}{S\arabic{table}}
\renewcommand{\thefigure}{S\arabic{figure}}
\renewcommand{\theHfigure}{S\arabic{figure}}
\renewcommand{\thelstlisting}{S\arabic{lstlisting}}
\renewcommand{\theHlstlisting}{S\arabic{lstlisting}}
\crefname{section}{Section}{Sections}

\begin{center}
{\Large\bfseries Supplementary material\par}
\end{center}
\vspace{4pt}
\noindent This supplementary material accompanies the paper. It contains the complete scenario oracle, coverage mappings, threat and invariant tables, the list of targeted tests, the full model-based, static-analysis and mutation results, cost and size tables, the claim schemas and the lifecycle diagram. Section, table, figure and equation numbers without the prefix S refer to the paper. All tables of results are generated from the raw outputs in the supplementary code archive by \texttt{paper/scripts/make\_tables.py}; the archive's README gives the commands that regenerate them.

\input{supplement/body}

\end{document}

%% file: preamble.tex
\setlist{nosep,leftmargin=*}
\newcolumntype{L}{>{\raggedright\arraybackslash}X}
\newcolumntype{P}[1]{>{\raggedright\arraybackslash}p{#1}}

\usepackage{hyperref}

\AtBeginEnvironment{table}{\linespread{1}\selectfont}
\AtBeginEnvironment{figure}{\linespread{1}\selectfont}
\AtBeginEnvironment{lstlisting}{\linespread{1}\selectfont}

\lstdefinelanguage{Solidity}{
  keywords={function,returns,return,if,else,for,external,internal,view,memory,calldata,storage,
    uint256,uint16,uint32,uint64,bytes32,bool,address,struct,contract,emit,revert,true,false},
  sensitive=true,
  comment=[l]{//},
  morestring=[b]"
}

\input{tables/macros}

\newcommand{\breakableunderscore}{\textunderscore\allowbreak}
\newcommand{\code}[1]{{\ttfamily\let\_\breakableunderscore #1}}

\crefformat{appendix}{#2#1#3}
\Crefformat{appendix}{#2#1#3}
\crefformat{subappendix}{#2#1#3}

%% file: tables/macros.tex
\newcommand{\OracleScenarios}{26}
\newcommand{\OracleSuppScenarios}{4}
\newcommand{\OracleBZeroAgree}{10}
\newcommand{\OracleBZeroApplicable}{23}
\newcommand{\OracleBZeroFalseAccept}{13}

\newcommand{\OracleBOneFalseAccept}{6}

\newcommand{\SuppBTwoAgree}{2}
\newcommand{\SuppBTwoApplicable}{4}

\newcommand{\SuppBTwoFAgree}{4}
\newcommand{\SuppBTwoFApplicable}{4}

\newcommand{\StageIssuance}{10}
\newcommand{\StageTransfer}{12}
\newcommand{\StageGovernance}{1}
\newcommand{\StageEvidence}{1}
\newcommand{\StageRedemption}{2}
\newcommand{\CondPositive}{4}
\newcommand{\CondBoundary}{3}
\newcommand{\CondNegative}{19}
\newcommand{\ModelSeeds}{20}
\newcommand{\ModelSteps}{500}

\newcommand{\ModelTransferTotalBTwoAdversarial}{2797}
\newcommand{\ModelTransferOKBTwoAdversarial}{78}
\newcommand{\ModelTransferOKPctBTwoAdversarial}{2.8}

\newcommand{\ModelTransferTotalBTwoMaintained}{3576}
\newcommand{\ModelTransferOKBTwoMaintained}{1283}
\newcommand{\ModelTransferOKPctBTwoMaintained}{35.9}

\newcommand{\ModelIssuanceDrawnTotalBTwoAdversarial}{1229}
\newcommand{\ModelIssuanceDrawnOKBTwoAdversarial}{4}

\newcommand{\ModelIssuanceDrawnTotalBTwoMaintained}{1449}
\newcommand{\ModelIssuanceDrawnOKBTwoMaintained}{108}

\newcommand{\RepTransferTotalBTwo}{360}
\newcommand{\RepTransferOKBTwo}{22}

\newcommand{\RepIssueDrawnTotalBTwo}{125}
\newcommand{\RepIssueDrawnOKBTwo}{3}

\newcommand{\ModelNonDebtBTwoAdversarial}{418}
\newcommand{\ModelNonDebtBTwoFAdversarial}{400}

\newcommand{\ModelNonDebtBTwoMaintained}{222}
\newcommand{\ModelNonDebtBTwoFMaintained}{119}
\newcommand{\GasTransferBZero}{50{,}492}

\newcommand{\GasIssueOneBZero}{59{,}772}

\newcommand{\GasTransferBOne}{78{,}234}

\newcommand{\GasTransferBTwo}{86{,}326}

\newcommand{\GasIssueThirtyTwoBTwo}{1{,}784{,}376}

\newcommand{\GasTransferBTwoF}{86{,}439}

\newcommand{\GasTransferDeltaOne}{27{,}742}
\newcommand{\GasTransferDeltaTwo}{35{,}834}

\newcommand{\GasTransferDeltaTwoOne}{8{,}092}

\newcommand{\OverheadTransferOne}{54.9}
\newcommand{\OverheadTransferTwo}{71.0}

\newcommand{\GasTransferRange}{12}

\newcommand{\GasRenewRights}{63{,}103}
\newcommand{\GasRenewClassification}{57{,}836}
\newcommand{\GasRenewStatus}{58{,}122}
\newcommand{\GasRenewValuation}{62{,}790}

\newcommand{\GasInlineDelta}{35{,}257}

\newcommand{\GasInlinePct}{57.3}
\newcommand{\DeployOverheadTwo}{3.0}

\newcommand{\IssueSlopeBOne}{39{,}295}

\newcommand{\IssueSlopeBTwo}{48{,}554}

\newcommand{\ScanDirect}{86{,}445}
\newcommand{\ScanFirst}{109{,}070}
\newcommand{\ScanFirstPos}{3}
\newcommand{\ScanLast}{458{,}523}
\newcommand{\ScanLastPos}{32}
\newcommand{\ScanSlope}{12{,}050}

\newcommand{\PriceLow}{1}
\newcommand{\PriceMid}{10}
\newcommand{\PriceHigh}{50}
\newcommand{\CyclesPerYear}{13}

\newcommand{\GasLiveness}{241{,}852}
\newcommand{\GasLivenessYear}{3{,}144{,}071}

\newcommand{\GasFullYearThirtyTwo}{77{,}187{,}019}

\newcommand{\CostTransferDeltaHigh}{0.001792}

\newcommand{\CostTransferDeltaLow}{0.000036}

\newcommand{\LatencyTrials}{200}
\newcommand{\LatencyTransferMedMin}{2.5}
\newcommand{\LatencyTransferMedMax}{3.4}
\newcommand{\CoverageLines}{99.0}
\newcommand{\CoverageLinesHit}{470}
\newcommand{\CoverageLinesFound}{475}
\newcommand{\SlitherInitial}{28}
\newcommand{\SlitherCurrent}{23}

\newcommand{\MutantsTotal}{6}

\newcommand{\TestsPassed}{82}
\newcommand{\TestsTS}{78}
\newcommand{\TestsSolidity}{4}
\newcommand{\TestsFailed}{0}
\newcommand{\FuzzRuns}{1{,}000}

\newcommand{\EnvHardhat}{3.11.1}
\newcommand{\EnvViem}{2.55.10}
\newcommand{\EnvOZ}{5.6.1}
\newcommand{\EnvSolc}{0.8.28}
\newcommand{\EnvRuns}{200}
\newcommand{\EnvEvm}{cancun}
\newcommand{\EnvHardfork}{osaka}

%% file: sections/abstract_preprint.tex
Security-token standards such as ERC-3643 decide who may hold a token, not what stands behind it. For a sukuk this gap matters: whether a certificate may trade at a negotiated price depends on what its portfolio contains, and a token that checks only investor identity keeps trading after every non-debt asset has been sold. We ask whether this tradability condition can be enforced on-chain, and at what cost. We encode the tradability conditions of the 2023 resolution of the Shariah Advisory Council of the Securities Commission Malaysia on Sukuk Wakalah bi al-Istithmar as issuance and transfer policies that read EIP-712 claims, signed by separate institutional roles, about each asset's rights, valuation, encumbrance and Shariah classification and about the allocation of proceeds. Across \OracleScenarios{} scenarios that we constructed on a synthetic instrument, an identity-only baseline wrongly allowed \OracleBZeroFalseAccept{} operations that the encoded policy denies, and asset evidence without the Shariah rule still allowed \OracleBOneFalseAccept. The full policy matched every expected outcome and added \GasTransferDeltaTwo{} gas (\OverheadTransferTwo\%) per transfer, independent of portfolio size. Our evaluation also exposes a trade-off that the constructed scenarios had missed. Checking the continuing condition through a single nominated asset keeps the cost of a transfer constant but can block trades the ruling allows, and a bounded fallback scan narrows this gap at about \ScanSlope{} gas per scanned asset. Our results indicate that Shariah tradability conditions can be enforced on-chain against attributable asset evidence, and that the way a ruling is encoded shapes which lawful trades a token permits. Github repo: \url{https://github.com/rischanlab/tokenized_sukuk}

%% file: sections/introduction.tex
\section{Introduction}
\label{sec:intro}

Tokenisation records ownership of a financial instrument on a programmable ledger and lets software enforce some of the conditions under which that ownership may change. The rights the token represents still rest on legal documents, custodians, valuations and institutions outside the ledger \citep{aldasoro2023continuum,iosco2025tokenization,sec2026tokenized}. For a sukuk this dependence is acute. A certificate represents an undivided interest in assets, usufructs or a portfolio, and whether it may be traded at a negotiated price depends on the composition of that portfolio and on the ruling of the competent Shariah authority \citep{aaoifi2018ss59,sac2023wakalah}.

Standards for regulated security tokens address a different question. ERC-3643, an Ethereum Request for Comments (ERC) standard for permissioned tokens, makes every transfer conditional on the receiver's verified identity and on a compliance contract, and gives an agent powers to pause, freeze and force transfers \citep{lebrun2021erc3643}. Its signed claims concern investors. The standard defines no representation of statements about the assets behind the token, so a token that implements only its identity checks and a compliance contract unaware of the portfolio continues to transfer after every non-debt asset in a sukuk portfolio has been sold.

This paper examines one rule: the resolution of 16 March 2023 of the Shariah Advisory Council (SAC) of the Securities Commission Malaysia on Sukuk Wakalah bi al-Istithmar, a structure in which an agent, the \emph{wakeel}, invests the proceeds of issuance in a portfolio of assets on behalf of the certificate holders. The resolution requires that at least 33\% of the proceeds of each issuance be invested in non-debt investment assets at the time of issuance, states that the percentage need not be maintained during the tenure, and requires that non-debt investment assets remain a component of the Wakalah portfolio at all times \citep{sac2023wakalah}. A token that checks only investor identity ignores both conditions; a token that enforces a permanent 33\% floor blocks transfers that the resolution permits. Both conditions depend on facts the ledger cannot observe, namely which assets the wakeel holds, what they are worth, whether they are encumbered and how a Shariah adviser classifies them. On-chain enforcement must therefore act on statements by identifiable parties, which raises the oracle problem \citep{hassan2023trust} and questions of role separation and accountability specific to sukuk \citep{kunhibava2021sukuk,busari2021application}.

Existing sukuk work offers an issuance prototype with a cost analysis \citep{khan2022tokenization} and conceptual designs for tracing financing to assets \citep{shaikh2019blockchained,moppratama2024property}. The closest technical design, RWA-PoB (real-world asset proof of backing), gates issuance and redemption of tokenised Treasury products on one aggregate, role-signed reserve snapshot \citep{mafrur2026pob}. Among the studies reviewed in \cref{sec:background}, none ties a cited sukuk tradability rule to signed, expiring evidence about individual assets and evaluates the result against a regulated-token baseline.

The study asks four questions. RQ1: how can signed claims about assets, rights and allocations be integrated into tokenised sukuk contracts so that decisions rest on attributable and current evidence, and what do such claims fail to establish? RQ2: how can one specified tradability rule be encoded as auditable issuance and transfer policies, with interpretation choices kept separate from the ruling? RQ3: how do the controls behave when asset composition, rights, valuations or Shariah evidence change, including the cases in which they deny operations the rule permits? RQ4: what gas, attestation and governance costs does the evidence layer add to a comparable regulated-token baseline?

The study follows a design-science approach built around a reproducible prototype. To separate the effect of asset evidence from that of the Shariah rule, it compares four configurations of the same token. B0 is a simplified baseline in the style of ERC-3643 that checks investor identity only. B1 adds signed evidence about assets and allocations without any Shariah rule. B2 adds the encoding of the SAC resolution and, to keep the cost of a transfer independent of portfolio size, checks the continuing condition through one nominated asset, called the \emph{witness}. B2F is B2 with a fallback that scans up to 32 allocated assets when the witness fails. The study makes three contributions.
\begin{enumerate}
\item An evidence model in which six institutional roles sign claims in the typed-data format of Ethereum Improvement Proposal (EIP) 712 about individual assets and about how the proceeds of each tranche, a separate issuance of certificates, are allocated, bound to one instrument, chain and registry, with one role per key, validity windows, sequence numbers, revocation, and encumbrance treated as negative evidence (\cref{sec:architecture}).
\item A versioned encoding of the SAC resolution that tests the 33\% minimum only at issuance and the continuing non-debt component at transfer, with each interpretation choice and each additional control identified as the authors' own (\cref{sec:policy}).
\item A matched evaluation of B0, B1, B2 and B2F against 26 author-constructed scenarios with expected outcomes and four supplementary scenarios, complemented by model-based differential testing, fuzzing, static analysis, mutation checks and gas measurements. The evaluation documents a case in which B2 denies transfers that the resolution permits, and the limited reach of the B2F correction (\cref{sec:evaluation}).
\end{enumerate}
The individual mechanisms have precedents; the contribution lies in combining asset-level evidence with an executable Shariah tradability policy and measuring the resulting controls, including their failure modes. \Cref{sec:background} reviews related work. \Cref{sec:design} specifies the policy and trust model and describes the design and its implementation. \Cref{sec:evaluation} sets out the evaluation and its results, \cref{sec:discussion} answers the four research questions and states the limitations, and \cref{sec:conclusion} concludes. \Cref{app:abbrev} lists the abbreviations and identifiers used in the paper.

%% file: sections/background.tex
\section{Background and related work}
\label{sec:background}

\subsection{Sukuk tradability rules}

Where a sukuk portfolio consists of receivables, a sale of the certificate is a sale of debt and is restricted accordingly; where it consists of tangible or other non-debt assets, the certificate can be traded at a negotiated price. Portfolios that mix both therefore need a composition rule, and standard setters differ on it. Shari'ah Standard No.~59 of the Accounting and Auditing Organization for Islamic Financial Institutions (AAOIFI) requires that the ratio of tangible and similar assets not fall below 33\% at any time and that a fall caused by a supervening event be corrected within a period set by the Shariah board \citep[cl.~8/2/2--8/2/3]{aaoifi2018ss59}. The SAC resolution of 16 March 2023 pairs an issuance test with a weaker continuing condition: at least 33\% of the proceeds of each issuance must be invested in non-debt investment assets at the time of issuance, the percentage need not be maintained throughout the tenure, and non-debt investment assets must at all times be a component of the Wakalah portfolio; under these conditions the sukuk may be traded at any negotiated price without being subject to bai' al-dayn, the rules that govern the sale of debt \citep{sac2023wakalah}. Standard IFSB-19 of the Islamic Financial Services Board adds a duty to disclose tradability limits identified by the Shariah adviser \citep{ifsb2017}. This paper encodes the SAC resolution only.

\subsection{Regulated security tokens and signed claims}

ERC-3643 requires for every transfer that the token is not paused, that neither wallet is frozen, that the receiver is verified in an identity registry, and that a compliance contract's \code{canTransfer} returns true; minting and forced transfers check only the receiver \citep{lebrun2021erc3643}. Verification rests on claims about the investor, signed by issuers in a trusted-issuers registry. The legal effect of a token transfer is settled by private law \citep{garciateruel2021digital}, and the rights of holders can differ with the way a security is tokenised \citep{sec2026tokenized}.

Contracts that act on off-chain facts depend on oracles, which reintroduce trusted parties \citep{hassan2023trust}. Proof-of-reserves schemes typically examine on-chain and off-chain assets separately \citep{lazirko2025proof}. RWA-PoB applies role-based attestation to tokenised Treasury products: five roles sign one EIP-712 snapshot of reserves, liabilities, liquidity and policy, an on-chain controller checks coverage and liquidity ratios before issuance and redemption, and redemption burns tokens into a pending obligation that a settlement role discharges \citep{mafrur2026pob}. That design authenticates institutional statements and, as its authors state, does not prove the existence of the assets. EIP-712 provides typed, domain-separated signatures and leaves replay protection to the application \citep{bloemen2017eip712}; ERC-1271 lets contract wallets validate signatures \citep{eip1271}. The claims used here borrow the issuer, subject, validity and status concepts of World Wide Web Consortium (W3C) Verifiable Credentials \citep{w3c2025vc} but are not W3C credentials.

\subsection{Sukuk on blockchain}

\citet{khan2022tokenization} coded the basic functions of a Sukuk al-Murabahah issuance in Solidity and reported gas per function. \citet{shaikh2019blockchained} proposed tracing sukuk financing to financed assets through a three-ledger model, and \citet{moppratama2024property} developed a stakeholder and data model for property sukuk in which Shariah auditors record compliance results; neither reports an implementation evaluated against a rule. Legal and Shariah reviews find that blockchain can make underlying assets and cash flows more transparent while raising legal, Shariah and cyber risks \citep{kunhibava2021sukuk}, that automation does not remove third parties \citep{busari2021application}, and that automated execution sits uneasily with the interpretive role of Shariah governance \citep{alsalih2025auditability}; reviews document few issuances and pilots \citep{pekerti2024smart,mousavi2025transforming}. A comparative review of tokenised sukuk and tokenised Treasury funds proposed, without implementation, a rules-and-evidence chain running from the holder's legal claim to exception handling \citep{mafrur2026preprint}. Table~S1 in the supplementary material compares these works feature by feature. The comparison covers the studies cited here and does not result from a systematic search. Within it, the present design is the only one that evaluates per-asset, role-specific evidence against a named tradability ruling and a matched regulated-token baseline.

%% file: sections/design.tex
\section{Design of the prototype}
\label{sec:design}

This section specifies the encoded policy and the trust model (\cref{sec:policy,sec:layers,sec:trust}), then describes the architecture, the issuance and transfer policies, governance and the implementation (\cref{sec:architecture,sec:issuancepolicy,sec:lifecycle,sec:implementation}).

\subsection{The encoded rule and its interpretation}
\label{sec:requirements}
\label{sec:policy}

The prototype implements one policy, \code{MY-SAC-WAKALAH-2023} version~1, for a synthetic Sukuk Wakalah bi al-Istithmar. The resolution has four elements \citep{sac2023wakalah}: (R1) at least 33\% of each issuance's proceeds is invested in non-debt investment assets, tested at the time of each issuance; (R2) the percentage need not be maintained during the tenure; (R3) non-debt investment assets remain a component of the Wakalah portfolio at all times; and (R4) under these conditions the sukuk may be traded at a negotiated price without being subject to bai' al-dayn. The encoding separates, for each element, what the resolution states, how the authors interpret it, what the contracts check and what is left unimplemented. \Cref{tab:trace} gives the full traceability, and this section summarises it.

\begin{table}[htbp]
\centering
\caption{Traceability from the SAC resolution to contract checks. Basis: SAC, stated in the resolution; Int., the authors' interpretation; Ctl., a control added by the authors. Scenario identifiers (S, G) and test identifiers (T) refer to the evaluation in \cref{sec:evaluation}; the scenarios are listed in Table~S5 and \cref{tab:supp}, the tests in Table~S8.}
\label{tab:trace}
\footnotesize
\setlength{\tabcolsep}{4pt}
\begin{tabularx}{\textwidth}{@{}P{2.5cm}LP{0.75cm}P{3.8cm}P{2.1cm}@{}}
\toprule
Element & Interpretation or control & Basis & Contract check & Tested by \\
\midrule
R1: at least 33\% of proceeds invested in non-debt assets, at each issuance & Tested at every tranche authorisation and at no other time & SAC & \code{authorizeIssuance}, \cref{eq:r1} & S01--S04, T01, T02 \\
 & ``At least 33\%'' read as 3,300 basis points, not one third & Int. & integer arithmetic & S02, S03 \\
 & Denominator is the tranche's proceeds; unallocated proceeds count in the denominator only & Int. & as above & S01 \\
 & ``Invested'' evidenced by a wakeel and trustee co-signed allocation; acquisitions not checked & Int. & tranche signatures & S07, S12, S13 \\
 & Non-debt status is a current Shariah adviser classification under the active policy version & Int. & \code{nonDebtEligible} & S04, T10 \\
 & Encumbered assets count neither as backing nor as non-debt & Ctl. & \code{backingEligible} & S14 \\
 & Allocation per asset bounded by its current attested value, within the instrument & Ctl. & \code{\_checkAllocation} & S08, S15, G04, T15, T19, T20 \\
\midrule
R2: percentage need not be maintained & No re-test of the share after issuance & SAC & none & S05, G03 \\
\midrule
R3: non-debt assets a component at all times & Satisfied when some asset is currently non-debt-eligible & Int. & -- & -- \\
 & B2 checks one nominated witness; B2F also scans up to 32 recorded candidates & Ctl. & \code{checkTransferEvidence}; B2F: fallback in \code{canTransfer} & S06, S11, S14, S18, S19, G01, G02, T21 \\
\midrule
R4: negotiated-price trading without bai' al-dayn when R1 and R3 hold & Rules on bai' al-dayn taken to apply when R3 fails (authors' inference); no price is observable on-chain, so no price rule is implemented & Int. & none & not tested \\
 & Ordinary transfers fail closed while R3 is not evidenced; no fail-open mode is implemented & Ctl. & \code{canTransfer} & S06 (blocked), S23 (allowed); no price involved \\
\midrule
Not in the resolution & Current Shariah status (endorsed) required for issuance and transfer & Ctl. & \code{authorizeIssuance}, \code{checkTransferEvidence} & S09, S24 \\
 & Role-bound, current, non-replayed claims; policy quorum; suspension; maturity; redemption & Ctl. & registry, engine, token & S07--S13, S16--S22, S25, S26 \\
\bottomrule
\end{tabularx}
\end{table}

\paragraph{R1, issuance minimum} For a tranche with proceeds $P$ and allocations $x_a$ to assets $a$, of which the set $N$ is currently classified as non-debt, the engine authorises issuance only if
\begin{equation}
\label{eq:r1}
10^{4} \sum_{a \in N} x_a \;\geq\; 3{,}300 \cdot P ,
\end{equation}
evaluated in integer arithmetic at each tranche authorisation and at no other time. Four interpretation choices enter here: ``at least 33\%'' is read as 33.00\%, not one third; the denominator is the tranche's proceeds, so unallocated proceeds count in the denominator only; the non-debt status of an asset is a current classification by the Shariah adviser under the active policy version; and ``invested'' is evidenced by an allocation co-signed by the wakeel and the trustee. The last choice is the weakest link: a co-signed allocation is a statement about the use of proceeds, not a record of executed acquisitions (\cref{sec:layers}). Two controls that the resolution does not address are added: encumbered assets count neither as backing nor as non-debt assets, and allocations to an asset are bounded, cumulatively within the instrument, by its current attested value.

\paragraph{R2 and R3, the continuing condition} The share is not re-tested after issuance. R3 is read as satisfied when at least one asset of the portfolio is currently \emph{non-debt-eligible}: it has a current rights claim, a current valuation with a positive value and, since B1 and B2 exclude encumbered assets, no effective encumbrance, which together make it \emph{backing-eligible}, and a current classification as non-debt under the active policy version. Configuration B2 checks this through one nominated \emph{witness} asset, so that the cost of a transfer does not depend on the size of the portfolio; B2F additionally scans up to 32 previously allocated assets, its \emph{candidates}, when the witness fails (\cref{sec:transferpolicy}). The witness is the authors' implementation device, and the cases in which it diverges from the reading of R3 are reported in \cref{sec:falsereject}.

\paragraph{R4, the price consequence} The resolution states what holds when R1 and R3 are met. The authors infer that, when the non-debt component is lost, trading becomes subject to the rules on bai' al-dayn; this inference, and what those rules would require of a secondary trade, need confirmation by a qualified Shariah adviser. A token transfer carries a sender, a receiver and an amount but no price, and the prototype contains no settlement or delivery-versus-payment module, so the contracts can neither observe a trade price nor enforce any price-dependent consequence. In place of such enforcement, B2 suspends ordinary transfers while the non-debt component is not evidenced. This fail-closed response is an engineering control chosen by the authors; the resolution does not require transfers to stop. The prototype has no fail-open mode: governance can only adopt a policy version that disables the non-debt rules, which replaces the non-debt test with the B1 backing test and also removes the issuance minimum. Two evaluation scenarios therefore concern R4 only through this control: S06, in which transfers are blocked once the non-debt component is lost, and S23, in which an ordinary transfer is allowed while R1 and R3 hold (\cref{sec:oracle}). Neither involves a price, and the R4 price treatment is untested.

\paragraph{Controls outside the resolution} B2 also requires a current Shariah status attestation (\emph{endorsed}) under the active policy version for issuance and transfer, and binds Shariah judgements to the policy version under which they were made, so that after a policy change they stop counting until re-attested; rights and valuation claims are not bound to a version. The remaining checks concern claim integrity, governance and lifecycle. The encoding is specific to this resolution; the AAOIFI position, with a continuing floor and a restoration period \citep{aaoifi2018ss59}, would require different logic.

\subsection{What a signed claim establishes}
\label{sec:layers}

The term ``proof of backing'' conflates five questions (\cref{tab:layers}). The contracts answer three: whether a statement was signed by the key registered for the required role (\emph{authenticity}), whether it is still in force (\emph{currency}), and whether the current statements jointly satisfy the policy (\emph{eligibility}). They do not establish whether the asset exists, is worth the attested value or received the proceeds (\emph{off-chain truth}), nor whether holders can enforce their rights (\emph{legal rights}). In this paper, \emph{asset evidence} means a set of current, role-attributed statements about an asset, not proof that the asset exists or backs the certificates. The R1 check uses the co-signed allocation as its evidence of investment; the prototype has no link to payments or acquisitions, so signed allocation is not independent evidence of executed investment.

\begin{table}[htbp]
\centering
\caption{Layers of assurance about asset evidence and what the prototype checks at each.}
\label{tab:layers}
\footnotesize
\begin{tabularx}{\textwidth}{@{}P{2.0cm}P{3.3cm}LP{3.4cm}@{}}
\toprule
Layer & Question & Checked by the prototype & Not established; relied on \\
\midrule
Authenticity & Was the statement signed by the key registered for this role? & EIP-712 digest over all fields; signer registered for the claim type; chain, registry and instrument binding; ERC-1271 for contract wallets & that the key holder is the named institution and acted honestly; attestor onboarding \\
Currency & Is the statement still in force? & validity window of at most 30 days; sequence numbers; claim and attestor revocation, evaluated at each use & that nothing changed off-chain within the window; timely attestor action \\
Eligibility & Do the current statements satisfy the policy? & backing and non-debt eligibility; issuance minimum; allocation within attested value within the instrument & sufficiency of assets against certificates outstanding; uniqueness of pledges across instruments \\
Off-chain truth & Does the asset exist, is it worth the attested value, were proceeds invested? & none; document hashes are signed and, for claims, emitted in events, not stored & existence, condition, valuation, disbursement; attestors, audit, supervision \\
Legal rights & Can holders enforce their claim, including in insolvency? & none & ownership, priority, true sale; offering documents and law \\
\bottomrule
\end{tabularx}
\end{table}

\subsection{Trust model, governance exceptions and residual risks}
\label{sec:trust}

Each attestor key holds exactly one role and is never reassigned after revocation, and each claim type can be signed only by its designated role: the legal verifier signs rights, the valuer valuations, the custodian encumbrance notices, the Shariah adviser classifications and the instrument's Shariah status, and the wakeel and trustee co-sign tranche allocations. The contracts cannot prevent one institution from operating keys in several roles; separating the institutions is a governance obligation that the design makes visible but does not enforce. Table~S3 in the supplementary material lists every party's on-chain authority, and Table~S2 lists the eight safety invariants tested: issuance only under the minimum and valid evidence (I1); no ordinary transfer without an evidenced non-debt component (I2); authorised and logged governance (I3); no update of authoritative state by revoked, expired, older or foreign claims (I4); no allocation beyond attested value or proceeds within the instrument (I5); consistent redemption accounting (I6); a pre-check on every balance change except three logged privileged paths (I7); and attributable suspensions and non-retroactive policy changes (I8). None has been proved formally.

Several powers are deliberate exceptions to decentralised control. Policy activation requires the approval of two of three governors and a one-day delay from the proposal, but attestor onboarding and revocation are held by one registry governor key, which therefore decides whose signatures count. The same key can revoke any stored claim, including an encumbrance notice, and, as administrator of the registry, can grant the governor role to further keys without a quorum. Any one governor can suspend the instrument, while resumption requires a quorum, so a single governor can halt ordinary trading. The agent's forced transfer bypasses the compliance contract and hence the Shariah rule, as in ERC-3643; it requires a verified receiver and a justification hash and emits an event, which makes its use attributable without preventing misuse. In the test deployment one key acts as agent, first engine governor, registry governor and registry administrator.

\paragraph{Residual risks}\label{sec:threats}The threat model (Table~S4) covers bypass of transfer rules, forged, altered, replayed or withheld claims, compromised attestor keys, out-of-order or contradictory evidence, double counting, unilateral policy changes and double redemption. Seven risks remain by construction: (a)~false content under a valid signature passes every check; (b)~an asset sold off-chain keeps counting until its claims expire or are revoked, a window bounded by the 30-day maximum validity; (c)~the wakeel and trustee together can authorise any allocation the attested values permit, and the valuer and legal verifier together can create those values; (d)~no rule compares the portfolio's attested value with the certificates outstanding, so transfers continue when that value collapses (supplementary scenario G03, \cref{sec:oracle}), because continuing coverage is outside the tradability condition encoded here; (e)~invariant I5 holds within one instrument, and a second instrument can allocate against the same asset undetected (test T15, \cref{sec:assurance_design}); (f)~forced transfers bypass the Shariah rule, one governor can suspend trading, and fail-closed checks turn missing evidence into blocked transfers that holders cannot remedy; and (g)~because the sequence counter is shared by all signers of a claim type for a subject, a compromised key that is not yet revoked can submit the largest possible sequence number and so prevent any later claim of that type for that subject.

\subsection{Architecture and evidence model}
\label{sec:architecture}

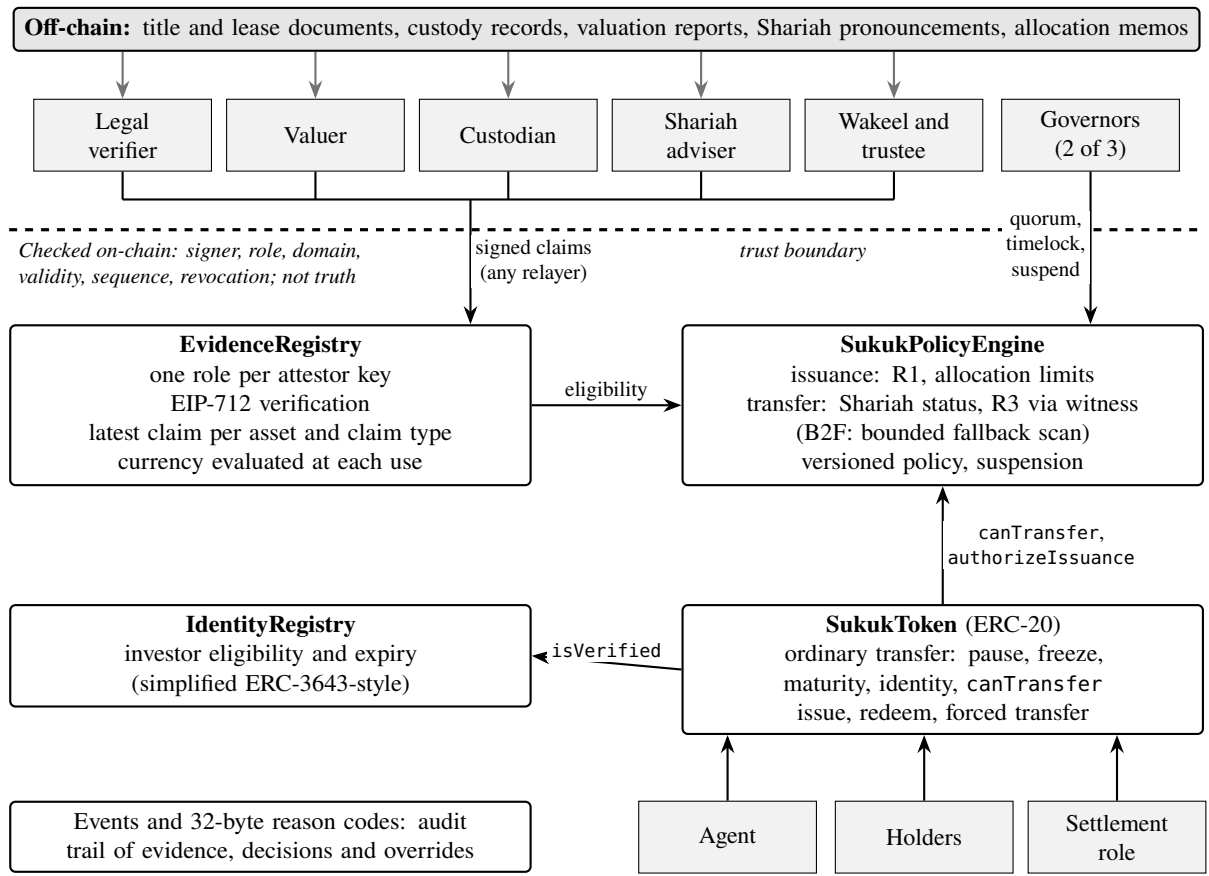
\begin{figure}[htbp]
\centering
\begin{tikzpicture}[
  font=\small,
  box/.style={draw, thick, rounded corners=2pt, align=center, inner sep=4pt},
  role/.style={draw, align=center, inner sep=3pt, minimum width=2.35cm, minimum height=0.95cm, fill=black!5},
  arr/.style={-{Stealth[length=2.4mm]}, thick},
  lab/.style={font=\footnotesize, align=center, fill=white, inner sep=1.5pt}
]
\node[box, text width=15.4cm, fill=black!10] (docs) at (0,7.2) {\textbf{Off-chain:} title and lease documents, custody records, valuation reports, Shariah pronouncements, allocation memos};
\node[role] (legal) at (-6.4,5.8) {Legal\\verifier};
\node[role] (valuer) at (-3.85,5.8) {Valuer};
\node[role] (cust) at (-1.3,5.8) {Custodian};
\node[role] (shariah) at (1.25,5.8) {Shariah\\adviser};
\node[role] (wt) at (3.8,5.8) {Wakeel and\\trustee};
\node[role] (gov) at (6.4,5.8) {Governors\\(2 of 3)};
\foreach \n in {legal,valuer,cust,shariah,wt} \draw[arr, black!55] (docs.south -| \n.north) -- (\n.north);
\draw[dashed, very thick] (-7.9,4.55) -- (7.9,4.55);
\node[font=\footnotesize\itshape, anchor=north] at (2.6,4.5) {trust boundary};
\node[font=\footnotesize\itshape, align=left, anchor=north west] at (-7.9,4.5) {Checked on-chain: signer, role, domain,\\validity, sequence, revocation; not truth};
\node[box, text width=6.6cm, anchor=north west] (reg) at (-7.9,3.3) {\textbf{EvidenceRegistry}\\one role per attestor key\\EIP-712 verification\\latest claim per asset and claim type\\currency evaluated at each use};
\node[box, text width=6.6cm, anchor=north east] (eng) at (7.9,3.3) {\textbf{SukukPolicyEngine}\\issuance: R1, allocation limits\\transfer: Shariah status, R3 via witness\\(B2F: bounded fallback scan)\\versioned policy, suspension};
\foreach \n in {legal,valuer,cust,shariah,wt} \draw[thick] (\n.south) -- (\n.south |- 0,4.95);
\draw[thick] (legal.south |- 0,4.95) -- (wt.south |- 0,4.95);
\draw[arr] (-1.8,4.95) -- node[lab, right] {signed claims\\(any relayer)} (-1.8,3.3);
\draw[arr] (reg.east) -- node[lab, above] {eligibility} (eng.west);
\draw[arr] (gov.south) -- node[lab, left] {quorum,\\timelock,\\suspend} (gov.south |- eng.north);
\node[box, text width=6.6cm, anchor=north east] (tok) at (7.9,-0.4) {\textbf{SukukToken} (ERC-20)\\ordinary transfer: pause, freeze,\\maturity, identity, \code{canTransfer}\\issue, redeem, forced transfer};
\node[box, text width=6.6cm, anchor=north west] (id) at (-7.9,-0.4) {\textbf{IdentityRegistry}\\investor eligibility and expiry\\(simplified ERC-3643-style)};
\draw[arr] (tok.north) -- node[lab, right] {\code{canTransfer},\\\code{authorizeIssuance}} (eng.south);
\draw[arr] (tok.west) -- node[lab, above] {\code{isVerified}} (id.east);
\node[role, anchor=north] (agent) at (1.6,-3.0) {Agent};
\node[role, anchor=north] (hold) at (4.2,-3.0) {Holders};
\node[role, anchor=north] (settle) at (6.75,-3.0) {Settlement\\role};
\draw[arr] (agent.north) -- (agent.north |- tok.south);
\draw[arr] (hold.north) -- (hold.north |- tok.south);
\draw[arr] (settle.north) -- (settle.north |- tok.south);
\node[box, text width=6.6cm, anchor=north west] at (-7.9,-3.0) {Events and 32-byte reason codes: audit trail of evidence, decisions and overrides};
\end{tikzpicture}
\caption{Components and trust boundary of the prototype. Configuration B0 replaces the registry and engine with a compliance contract that allows every operation; B1 and B2 use the same contracts with different policy parameters; B2F replaces the engine with a variant that adds a fallback scan.}
\label{fig:architecture}
\end{figure}

\Cref{fig:architecture} shows the components and the trust boundary. Off-chain parties sign statements about documents they hold; the ledger verifies the signatures, the signers' roles and the statements' currency, and has no access to the documents. Three contracts carry the design: an \emph{evidence registry} that verifies and stores claims, a \emph{policy engine} that turns current claims into issuance and transfer decisions under a versioned policy, and a permissioned \emph{token} that consults the engine and an identity registry on every ordinary transfer and on issuance. Signatures are verified once, when a claim is submitted, and decisions read the stored result. Because a claim's validity rests on its signature, any account may submit it; such an account is called a \emph{relayer}.

\paragraph{Evidence model}\label{sec:evidence} 
Every claim is an EIP-712 \code{Attestation} with eleven fields: the instrument identifier, a claim type, a subject (an asset, or the instrument for Shariah status), a code, a flag, an amount, a policy version, the hash of the referenced off-chain document, a sequence number and a validity window (Listing~S1). The domain binds the signature to one chain and one registry contract, and the instrument identifier binds it to one sukuk. Five claim types are defined, each signable by exactly one role: rights, valuation, classification, encumbrance and Shariah status. A sixth structure, the \code{TrancheAllocation}, lists the assets to which a tranche's proceeds are allocated and must be co-signed by an active wakeel and an active trustee.

A claim is accepted only if it names this instrument, its signer is an active attestor holding the required role, its window has started, has not ended and does not exceed the registry's maximum validity, 30 days in the prototype, its sequence number exceeds the last one accepted for the same subject and type, and its signature verifies through OpenZeppelin's \code{SignatureChecker}, which accepts externally owned accounts and ERC-1271 contract wallets \citep{openzeppelin561,eip1271}. Sequence numbers decide which claim is authoritative, so a delayed or reordered submission cannot overwrite a newer one. Currency is evaluated at each use: a claim counts only if it has not been revoked or expired and its signer is still active. Revoking an attestor therefore invalidates every claim it has signed other than encumbrance notices, including claims made before a compromise, because the registry cannot tell which earlier signatures were honest. Encumbrance is negative evidence: a custodian's notice remains effective after it expires and after its signer is revoked, and ends only when superseded or revoked. When a rights or encumbrance claim is submitted and a current unencumbered rights claim then coexists with an encumbrance notice, the registry emits an \code{EvidenceConflict} event. Under a policy that excludes encumbered assets, as B1 and B2 do, an asset stops counting while an encumbrance notice is in force, with or without a conflict.

Verifying a bundle of signed claims inside each transfer was considered and rejected: it is more expensive (\cref{sec:cost}) and weaker, since an inline bundle cannot reveal a later revocation, a newer claim or an encumbrance notice without consulting the same stored state.

\subsection{Issuance and transfer policies}
\label{sec:issuancepolicy}

\paragraph{Issuance} 
The agent initiates issuance by passing the co-signed allocation and the recipients to the token, which asks the engine to authorise the total. The engine completes every check before writing state, so a denial leaves it unchanged. It rejects a suspended instrument, a reused tranche identifier, an invalid or expired co-signature, proceeds that differ from the amount to be minted, a tranche signed for another policy version and, under B2, a missing or lapsed Shariah status. Asset identifiers in the allocation must be strictly increasing, which excludes duplicates; each asset must be backing-eligible; its cumulative allocation may not exceed its current attested value; and the allocation total may not exceed the proceeds. Under B2 the engine then applies \cref{eq:r1}. On success it records the tranche with the policy version in force and, if the current witness is not eligible, nominates the first eligible asset of the tranche.

\paragraph{Transfer and the witness asset}\label{sec:transferpolicy} 
The transfer check makes one external call into the registry, which tests the Shariah status and then the witness. Under B2 the witness must be non-debt-eligible; under B1 it must be backing-eligible. The check therefore costs the same whatever the size of the portfolio. The price is \emph{liveness}, the ability of a transfer that the policy permits to execute without further intervention. If the witness loses eligibility while another non-debt asset still qualifies, R3 holds, yet B2 denies transfers until someone calls \code{setWitness} with a qualifying asset. Anyone may call it, because the engine accepts only an asset that satisfies the active policy, so an unprivileged relayer can restore transfers; until then, B2 falsely rejects transfers in the policy sense.

\paragraph{The fallback variant} Configuration B2F tests one correction. Its engine records each asset when the instrument first allocates to it, up to \code{MAX\_CANDIDATES}~$=32$. When the witness check fails with a component or evidence reason, \code{canTransfer} scans the recorded candidates in allocation order and allows the transfer if any is non-debt-eligible. The scan is a read-only view and does not change the stored witness, so the constant-cost path returns once a relayer re-nominates. The cap bounds the worst-case cost of a transfer, and assets allocated after the list is full are never scanned. Automatic re-nomination inside the transfer and sets of nominated witnesses were not implemented.

\subsection{Governance and lifecycle}
\label{sec:lifecycle}

Policy parameters (version, minimum non-debt share, which rules apply, whether encumbered assets are excluded and the hash of the policy document) change only through a proposal for the next version that at least two of the three governors approve and that any governor can activate no earlier than one day after the proposal was made; quorum and delay are deployment parameters. Activation does not alter recorded tranches, and because classifications and Shariah status are bound to a version, a new version that keeps the Shariah rules enabled fails closed until the Shariah adviser re-attests. The engine moves from \emph{Draft} to \emph{Active} with the first authorised tranche; any governor can suspend it, and resumption requires a quorum within the same suspension epoch. After maturity, a timestamp in the token, ordinary transfers and issuance stop and holders may request redemption, which burns their units into a pending claim that the settlement role discharges by recording payments; this burn-to-pending pattern follows RWA-PoB \citep{mafrur2026pob}. Figure~S1 in the supplementary material shows the state machine. The amount payable per unit is set off-chain by the dissolution terms.

\subsection{Implementation}
\label{sec:implementation}

The prototype is written in Solidity \EnvSolc{} with the \code{\EnvEvm} target and the optimiser enabled at \EnvRuns{} runs, and builds on OpenZeppelin Contracts \EnvOZ{} for ERC-20, access control, pausing, EIP-712 hashing and signature checking \citep{openzeppelin561}. Tests and experiments run on Hardhat \EnvHardhat{} with viem \EnvViem{} and the Node.js test runner \citep{hardhat3}. The production contracts are \code{SukukToken}, \code{IdentityRegistry} (a simplified stand-in for ERC-3643's identity registry), \code{BasicCompliance} (B0), \code{EvidenceRegistry}, \code{SukukPolicyEngine} (B1, B2) and \code{SukukPolicyEngineFallback} (B2F); Tables~S15 and S16 list them and their bytecode sizes.

\paragraph{One path for every balance change} The token overrides OpenZeppelin's internal \code{\_update} function, through which every mint, burn and transfer passes, including \code{transferFrom}. Unless a transaction-scoped flag is set, \code{\_update} runs the same pre-check that the public \code{preCheckTransfer} view exposes and reverts with \code{TransferNotCompliant(reason)} if it fails. The pre-check tests, in order, whether the token is paused, whether either wallet is frozen, whether the instrument has matured, whether the receiver is verified, and then calls the compliance contract. The flag is a Solidity \code{transient} variable that lives only for the current transaction, and only three privileged paths set it: issuance after authorisation, redemption burns and the agent's forced transfer. Each performs its own checks and emits an event.

\paragraph{Issuance as a policy decision} In ERC-3643 minting bypasses the compliance contract and checks only the receiver \citep{lebrun2021erc3643}. Here the token passes the tranche evidence to \code{authorizeIssuance}, mints only if the engine returns success, and still requires every recipient to be verified. In B0 the compliance contract allows every issuance, which reproduces the ERC-3643 behaviour.

\paragraph{Reason codes and arithmetic} Every denial of a transfer or an issuance authorisation carries a 32-byte ASCII reason, for example \code{NON\_DEBT\_COMPONENT\_UNVERIFIED}. The engine returns reasons without reverting, so the token can expose the same decision through a view and through revert data, and tests compare reasons as well as outcomes. Errors in claim submission, governance and token arguments revert with custom errors instead. The issuance minimum is evaluated as a cross-multiplication in basis points, which avoids division and rounding (\cref{lst:ratio}).

\begin{lstlisting}[caption={Issuance minimum and transfer check in \code{SukukPolicyEngine} (abridged).},label={lst:ratio}]
// authorizeIssuance, after signature, amount, version and allocation checks
if (p.applyNonDebtRules &&
    nonDebtSum * BPS < uint256(p.minNonDebtBps) * t.proceeds)
    return ReasonCodes.NON_DEBT_RATIO_BELOW_MINIMUM;

// canTransfer: one external call, independent of the number of assets
if (state == State.Suspended) return ReasonCodes.INSTRUMENT_SUSPENDED;
return evidence.checkTransferEvidence(witnessAssetId, p.version,
    p.applyNonDebtRules, p.requireShariahStatus, p.excludeEncumbered);
\end{lstlisting}

\paragraph{Storage} The registry keeps, per subject and claim type, the signer, a 40-bit expiry, the policy version, the code and two flags in one 256-bit slot, and stores valuation amounts separately for the allocation check. The engine keeps the cumulative allocation per asset and one record per tranche. Because a key holds only one role, the wakeel and trustee signatures necessarily come from different keys. An internal hook, called for each allocated asset with a flag that marks its first allocation, lets B2F record candidates; it is empty in \code{SukukPolicyEngine}.

%% file: sections/evaluation.tex
\section{Evaluation}
\label{sec:evaluation}

The evaluation compares the four configurations on one synthetic instrument (\cref{sec:experiments}) with a scenario oracle, model-based differential testing, targeted tests, static analysis, mutation checks and gas measurements (\cref{sec:oracle,sec:assurance_design}). \Cref{sec:results,sec:falsereject,sec:mbresults,sec:assurance,sec:cost} report the results.

\subsection{Configurations and synthetic instrument}
\label{sec:experiments}

All configurations deploy the same token and identity registry and are driven by the same test code; they differ only in the compliance contract and its parameters (\cref{tab:configs}). B0 is a simplified ERC-3643-style research baseline: it shares with ERC-3643 the receiver identity check on every transfer and mint, pause, wallet freeze and forced transfer, and omits on-chain identity contracts, claim topics, trusted-issuer registries, partial freezes, wallet recovery and modular compliance. It reproduces no deployed issuer's code, so overheads measured against it are overheads against this baseline only. B1 adds the evidence layer, tranche co-signatures, the allocation checks and the governance and lifecycle controls, without Shariah rules. B2 enables the issuance minimum, the continuing-component check and the Shariah status requirement. B2F is B2 with the fallback scan.

\begin{table}[htbp]
\centering
\caption{Experimental configurations. All share \code{SukukToken} and \code{IdentityRegistry}.}
\label{tab:configs}
\footnotesize
\begin{tabularx}{\textwidth}{@{}LP{1.9cm}P{1.9cm}P{1.9cm}P{1.9cm}@{}}
\toprule
Feature & B0 & B1 & B2 & B2F \\
\midrule
Compliance contract & basic (always allows) & policy engine & policy engine & engine with fallback \\
Receiver identity check on transfer, \code{transferFrom}, mint and forced transfer & yes & yes & yes & yes \\
Pause, wallet freeze, forced transfer with justification & yes & yes & yes & yes \\
Partial freeze, wallet recovery, on-chain identity claims & no & no & no & no \\
Signed rights, valuation and encumbrance evidence; tranche co-signatures & no & yes & yes & yes \\
Allocation within attested value; no double counting within the instrument & no & yes & yes & yes \\
Versioned policy, quorum, timelock, suspension & no & yes & yes & yes \\
Classification and Shariah status evidence & no & no & yes & yes \\
Issuance minimum (\code{minNonDebtBps}) & -- & 0 & 3,300 & 3,300 \\
Continuing non-debt component on transfer & no & backing witness only & witness & witness, then up to 32 candidates \\
\bottomrule
\end{tabularx}
\end{table}

The synthetic instrument is a one-year Sukuk Wakalah with three assets: a beneficial interest in a leased commercial property (ijarah, non-debt, 25 million units), a musharakah participation (non-debt, 15 million) and a commodity murabahah receivable (debt, 80 million). The first tranche raises 100 million units, allocates 25, 10 and 65 million to the three assets, a non-debt share of 35\%, and is issued to two investors. Mock documents stand in for the off-chain evidence whose hashes the claims carry; no real issuer, asset or investor is represented. One key is registered per attestor role.

\subsection{Scenario oracle}
\label{sec:oracle}

The scenario oracle is a set of scenarios, each with the outcome that every configuration is expected to produce. The primary oracle consists of \OracleScenarios{} scenarios, S01--S26. The authors constructed them around the resolution elements, the invariants and the threats of \cref{sec:design}, choosing for each element at least one case that violates it, cases on both sides of the 33\% threshold and a case at the inclusive expiry bound of the evidence. The number 26 is the outcome of this procedure; it was not derived statistically, and the scenarios are a targeted suite, not a sample of instrument states. The set contains \StageIssuance{} issuance, \StageTransfer{} transfer, \StageEvidence{} evidence-submission, \StageGovernance{} governance and \StageRedemption{} redemption scenarios, of which \CondPositive{} are positive controls, \CondNegative{} negative cases and \CondBoundary{} boundary cases. Every invariant and threat is exercised by at least one primary scenario; the R4 price treatment is not exercised, and S06 and S23 test only the authors' transfer control that stands in for it (\cref{sec:policy}). The full scenarios, their mapping to elements and the per-scenario decisions are in Tables~S5, S6 and S7 of the supplementary material. Each scenario starts from a fresh deployment, applies a setup for one fault condition and performs one operation, so combinations of simultaneous faults are left to the model-based runs.

Four supplementary scenarios, G01--G04 (\cref{tab:supp}), cover cases the primary set omits: the witness asset valued at zero (G01) or reclassified as debt (G02) while another non-debt asset remains eligible and no relayer re-nominates; all attested values falling to 1\% of their issuance values (G03); and a new allocation to an asset whose value was cut below the amount already allocated to it (G04). They were written after the B2 false rejection was known and are reported separately.

For every scenario the oracle records a \emph{design expectation}, what each configuration is specified to do, and a \emph{policy expectation}, what the authors' encoding requires. A false accept is an operation a configuration allows and the policy denies, and a false reject the converse. The oracle thus measures decision conformance to the specified policy. It is distinct from whether the portfolio actually satisfies the rule, whether the current claims establish that it does, and liveness (\cref{sec:transferpolicy}); the witness false rejection is a failure of liveness in which the rule holds and the evidence for the witness does not. The same authors wrote the specification, the expectations, the reference model and the contracts, so this separation can expose implementation errors but not a misreading of the resolution shared by all four.

\subsection{Model-based testing, targeted tests and cost measurement}
\label{sec:assurance_design}

\paragraph{Model-based differential testing}
A reference model of the policy, written in TypeScript from the specification by the same authors, and the contracts were driven by the same seeded random walk (mulberry32 generator). Each step draws a transfer (to an investor or to an unregistered wallet), an evidence event (valuation, re-attestation, custodian notice, revocation or Shariah status change), a time advance, a witness re-nomination, a suspension or resume vote, a tranche issuance, or a redemption or settlement step. After every step the test asserts that the contract's outcome equals the model's, comparing decision and reason code for transfers and issuances and success or failure for revocations, witness nominations, resume votes, redemptions and settlements, and that the invariants hold: issued units equal supply plus pending plus settled redemptions (I6), balances and allocations equal the model's (I5, I7), no unregistered wallet holds units (I7), the witness equals the model's and, under B2F, so does the number of recorded candidates. At each authorised B2 and B2F issuance the test also checks that the model's non-debt allocation met the minimum (I1). The \emph{adversarial} profile draws validity windows of one to ten days, zero valuations in a quarter of updates and frequent revocations, encumbrances and suspensions, so most operations are denied. The \emph{maintained} profile draws windows of ten to thirty days and rare disruptions, so successful transfers and issuances are common. B1, B2 and B2F were each run for \ModelSeeds{} seeds of \ModelSteps{} steps in each profile.

\paragraph{Targeted tests and additional assurance}
Targeted tests T01--T21 (Table~S8) check individual properties. Among them, T15 pledges one asset in two instruments, T16 cuts all valuations after issuance, T17 renews only the four claims the transfer check reads (the witness's rights, valuation and classification and the Shariah status) and lets the others lapse, T18 exercises the forced-transfer bypass, T19 and T20 test the exact bounds of the allocation checks, and T21 tests the fallback cap: 33 assets are allocated, 32 of them non-debt, every non-debt asset except the 33rd loses eligibility, and the transfer decision is recorded under B2 and B2F. Every transfer through the test helper asserts that the read-only pre-check and the executed transaction agree. Four Solidity fuzz properties were run at \FuzzRuns{} runs each with a fixed seed: under B2, for random allocations to two assets that never together exceed the proceeds, issuance is authorised exactly when each allocation is within its attested value and the non-debt share reaches 3,300 basis points (I1, I5); redemption accounting is conserved under arbitrary request and settlement sequences (I6); a claim is accepted exactly when its window contains the current time and is at most 30 days long (I4); and sequence numbers only increase (I4). Slither 0.11.6 \citep{feist2019slither} was run with all detectors on each deployable contract. Six hand-written mutants, each weakening one boundary, were run against the fuzz suite and the functional suites; a mutant counts as killed if at least one test fails \citep{jia2011mutation}.

\paragraph{Cost measurement}
The benchmark deploys each configuration, publishes the fixture's evidence with 30-day validity and issues the first tranche. It records gas for deployment; 1,000 transfers of one unit between holders with non-zero balances; 1,000 valuation updates; 100 renewals each of rights, classification and Shariah status; issuances allocating to 1, 2, 4, 8, 16 and 32 freshly evidenced assets; lifecycle actions; and, for B2F, transfers in which the first eligible candidate sits at each position of a full 32-entry list. It also compares the stored-state transfer check with an inline check that verifies four signed claims. Gas is read from transaction receipts and includes the 21,000-gas base cost. Fees in native-token units follow as $\text{fee} = \text{gas} \times p \times 10^{-9}$ for a gas price $p$ in gwei; the prices \PriceLow, \PriceMid{} and \PriceHigh{} gwei are assumptions, not observed market prices. All measurements were taken on Hardhat's in-process simulated network (EDR, \code{\EnvHardfork} rules, automine); line coverage was collected in a separate instrumented run, because instrumentation changes gas.

\subsection{Decisions on the scenario oracle}
\label{sec:results}

All \TestsPassed{} tests (\TestsTS{} TypeScript tests and \TestsSolidity{} Solidity fuzz properties) passed in the reported run, with \TestsFailed{} failures. Every measured count, gas figure and timing in this and the following subsections is generated from the raw output files by a script in the supplementary code.

All four configurations behaved as designed in every primary and supplementary scenario. \Cref{tab:decisions} compares their decisions with the policy expectation. Within the \OracleScenarios{} primary scenarios, B2 and B2F agreed with the policy in all \OracleScenarios, with no false accepts or false rejects. In the boundary scenarios an allocation of exactly 33.00\% of the proceeds was authorised and one of 32.99\% was denied (S02, S03), and in the negative control S05 the non-debt share fell to about 6\% after issuance while transfers continued, as R2 permits.

\begin{table}[htbp]
\centering
\caption{Agreement with the policy expectation. Primary: S01--S26; supplementary: G01--G04. FA, false accepts; FR, false rejects. Every configuration made its designed decision in every scenario. Per-scenario decisions are in Table~S5 and \cref{tab:supp}.}
\label{tab:decisions}
\small
\begin{tabular}{@{}lccccccc@{}}
\toprule
 & \multicolumn{3}{c}{Primary scenarios} & & \multicolumn{3}{c}{Supplementary scenarios} \\
\cmidrule(lr){2-4}\cmidrule(l){6-8}
Configuration & Agree & FA & FR & & Agree & FA & FR \\
\midrule
\input{tables/decision_compact.tex} 
\bottomrule
\end{tabular}
\end{table}

B0 agreed with the policy in \OracleBZeroAgree{} of \OracleBZeroApplicable{} applicable scenarios (the governance, evidence-ordering and suspension scenarios S17, S20 and S25 do not apply to it) and accepted \OracleBZeroFalseAccept{} operations that the policy denies; it rejected only what its identity, maturity and redemption checks cover. B1 accepted \OracleBOneFalseAccept{} such operations, all of which depend on Shariah evidence: the issuance minimum (S03, S04), the continuing component (S06, S11, S14) and the Shariah status (S24). The evidence layer alone thus closed the baseline's gaps concerning signer credentials (S07), chain binding (S13), replay (S12), expiry (S09), allocation limits (S08, S15) and key compromise (S18), and the Shariah rules closed the remainder. In the compromise scenario, revoking the legal verifier's key made every rights claim it had signed non-current, and transfers stayed denied until a new key had re-attested and the witness had been re-nominated (S18, S19).

\subsection{The witness false rejection and the fallback boundary}
\label{sec:falsereject}

When the witness asset was valued at zero while the musharakah asset remained non-debt-eligible (G01), B2 denied the transfer although R3 held; when the witness was reclassified as debt (G02), B2 again denied it. These are false rejections in the policy sense, caused by the single-witness design (B1 also denied G01, because its witness had lost its backing eligibility). B2F permitted both transfers through its scan, and test T03-F confirms that the scan leaves the stored witness unchanged and that B2F denies transfers once no recorded candidate is non-debt-eligible. Across the supplementary set B2 agreed with the policy in \SuppBTwoAgree{} of \SuppBTwoApplicable{} scenarios and B2F in \SuppBTwoFAgree{} of \SuppBTwoFApplicable{} (\cref{tab:supp}).

\begin{table}[htbp]
\centering
\caption{Supplementary scenarios, reported separately from the primary oracle. Policy: outcome required by the authors' encoding. A, allowed; D, denied. $^{\ast}$ marks a decision that differs from the policy. The last column gives B2's reason code.}
\label{tab:supp}
\footnotesize
\setlength{\tabcolsep}{3pt}
\begin{tabularx}{\textwidth}{@{}p{0.6cm}p{1.35cm}Lcccccp{3.2cm}@{}}
\toprule
ID & Operation & Scenario & Policy & B0 & B1 & B2 & B2F & B2 reason \\
\midrule
\input{tables/oracle_supp.tex} 
\bottomrule
\end{tabularx}
\end{table}

The B2F correction is bounded. In test T21 the instrument allocated to 33 assets, 32 of them classified as non-debt, so the candidate list held the first 32 allocated and omitted the 33rd. When every other non-debt asset lost eligibility, B2F denied the transfer with the same reason as B2 although a non-debt component remained; re-nominating the 33rd asset as witness restored transfers under both configurations. B2F therefore removes the false rejection only when an eligible asset lies within the first 32 allocated.

Three further tests document the scope of the controls. With every attested value cut to 1\% of its issuance value, all configurations continued to permit transfers (G03, T16), because no rule compares the portfolio's value with the certificates outstanding; a new allocation to the devalued ijarah asset was rejected under B1, B2 and B2F (G04). A second instrument with its own registry and engine allocated 25 million units against the ijarah asset that the first had already allocated in full (T15). The agent's forced transfer succeeded while the continuing-component check was failing and emitted an event naming the agent and the justification hash (T18).

\subsection{Model-based differential testing}
\label{sec:mbresults}

Across \ModelSeeds{} seeds of \ModelSteps{} steps in each profile for each of B1, B2 and B2F, the contracts' outcomes equalled the reference model's at every step, including the reason code of every transfer and issuance, and no invariant check failed. In the adversarial profile denials dominated: under B2, \ModelTransferOKBTwoAdversarial{} of \ModelTransferTotalBTwoAdversarial{} transfers (\ModelTransferOKPctBTwoAdversarial\%) succeeded and \ModelIssuanceDrawnOKBTwoAdversarial{} of \ModelIssuanceDrawnTotalBTwoAdversarial{} randomly drawn issuances were authorised; the fixed first tranche of each walk is excluded from these issuance figures. The maintained profile exercised the successful paths: \ModelTransferOKBTwoMaintained{} of \ModelTransferTotalBTwoMaintained{} transfers (\ModelTransferOKPctBTwoMaintained\%) succeeded and \ModelIssuanceDrawnOKBTwoMaintained{} of \ModelIssuanceDrawnTotalBTwoMaintained{} drawn issuances were authorised, and redemption requests and settlement confirmations were submitted in every configuration. Under the same seeds and step mix, B2F reduced the transfers denied for an unverified non-debt component from \ModelNonDebtBTwoMaintained{} to \ModelNonDebtBTwoFMaintained{} in the maintained profile and from \ModelNonDebtBTwoAdversarial{} to \ModelNonDebtBTwoFAdversarial{} in the adversarial profile; in this run its issuance outcomes were identical to B2's in every seed. The full outcome distribution is in Table~S9.

\subsection{Fuzzing, static analysis and mutation checks}
\label{sec:assurance}

All four fuzz properties held over \FuzzRuns{} runs each. Slither initially reported \SlitherInitial{} distinct findings; five medium-impact findings (local accumulators relying on implicit zero initialisation) and two low-impact findings (missing zero-address checks in two contracts) were fixed. The current run reports \SlitherCurrent{} findings, all of low impact, which concern comparisons with \code{block.timestamp} and external calls inside bounded loops and were retained by design; two of the loop findings concern the fallback scan of B2F, which the initial run did not analyse (Table~S11). Two mutants that loosened the allocation bounds initially survived both suites, which had tested over-allocation only with large margins; after T19 and T20 were added, every one of the \MutantsTotal{} mutants was killed by at least one suite, and the settlement mutant only by a fuzz property (Table~S12). Six mutants test selected boundaries and are not a measure of overall test adequacy. Line coverage of the production contracts was \CoverageLines\% (\CoverageLinesHit{} of \CoverageLinesFound{} lines); the unexecuted lines are input-validation and access-control branches.

\subsection{Cost}
\label{sec:cost}

A transfer between holders with existing balances cost \GasTransferBZero{} gas under B0, \GasTransferBOne{} under B1, \GasTransferBTwo{} under B2 and \GasTransferBTwoF{} under B2F with a valid witness (\cref{tab:gas}). Relative to B0, the evidence check added \GasTransferDeltaOne{} gas (\OverheadTransferOne\%) under B1 and \GasTransferDeltaTwo{} gas (\OverheadTransferTwo\%) under B2, and the difference between B2 and B1, \GasTransferDeltaTwoOne{} gas, reflects the additional Shariah status and classification checks. Because the transfer check reads one witness, its cost does not depend on the number of assets by construction; the benchmark measured it with three assets, and under B2F with a full 32-entry candidate list a transfer with a valid witness cost \ScanDirect{} gas. Verifying four signed claims inside the transfer instead of reading stored state would have added \GasInlineDelta{} gas (\GasInlinePct\%) to the check.

\begin{table}[htbp]
\centering
\begin{threeparttable}
\caption{Gas used per operation (transaction receipts, including the 21,000-gas base cost). Overheads are relative to the simplified ERC-3643-style baseline B0.}
\label{tab:gas}
\footnotesize
\setlength{\tabcolsep}{3.5pt}
\begin{tabular}{@{}lrrrrrr@{}}
\toprule
Operation & B0 & B1 & B2 & B2F & B1 vs B0 & B2 vs B0 \\
\midrule
\input{tables/gas.tex} 
\bottomrule
\end{tabular}
\begin{tablenotes}\footnotesize
\item[a] Role grants, identity registrations and attestor registrations.
\item[b] Holders with non-zero balances; the range across 1,000 transfers was at most \GasTransferRange{} gas in each configuration.
\item B0 ignores the allocation, so its issuance cost does not depend on the number of assets. Renewals refer to a claim slot that already holds a claim.
\end{tablenotes}
\end{threeparttable}
\end{table}

\begin{figure}[htbp]
\centering
\begin{tikzpicture}
\begin{axis}[
  name=left, width=0.49\textwidth, height=5.4cm,
  xmode=log, log basis x=2, xtick={1,2,4,8,16,32}, xticklabels={1,2,4,8,16,32},
  xlabel={Assets allocated in the tranche}, ylabel={Gas (thousands)},
  ymin=0, legend pos=north west, legend style={font=\scriptsize, draw=none, fill=none},
  tick label style={font=\scriptsize}, label style={font=\footnotesize},
  mark size=1.8pt, grid=major, grid style={black!10}]
\pgfplotsset{cycle list={{black, mark=square*},{black!55, mark=triangle*},{black, dashed, mark=*},{black!55, dotted, mark=o}}}
\input{tables/fig_issuance}
\legend{B0, B1, B2, B2F}
\end{axis}
\begin{axis}[
  at={(left.east)}, anchor=west, xshift=1.4cm,
  width=0.49\textwidth, height=5.4cm,
  xlabel={Position of first eligible candidate}, ylabel={Transfer gas (thousands)},
  xmin=0, xmax=33, ymin=0,
  tick label style={font=\scriptsize}, label style={font=\footnotesize},
  mark size=1.2pt, grid=major, grid style={black!10}]
\pgfplotsset{cycle list={{black, mark=*}}}
\input{tables/fig_scan}
\end{axis}
\end{tikzpicture}
\caption{Left: issuance gas against the number of freshly evidenced assets allocated (one recipient; logarithmic x-axis; linear growth under B1 and B2). Right: B2F transfer gas when the witness has failed, against the position of the first non-debt-eligible candidate in the 32-entry list; positions 1 and 2 hold the failed witness and the debt-classified asset.}
\label{fig:costs}
\end{figure}
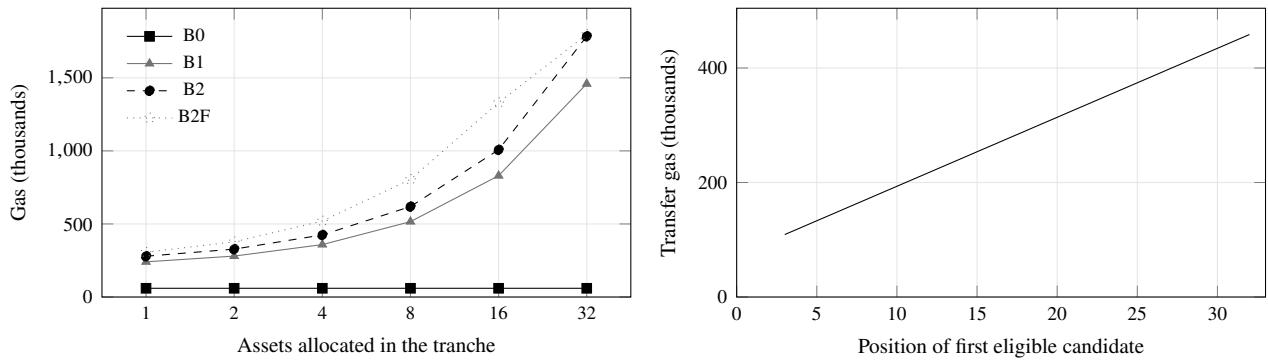

Issuance is where the evidence layer is expensive. Under B1 and B2 issuance gas grew linearly with the number of freshly evidenced assets (\cref{fig:costs}, left): a least-squares fit over 1 to 32 assets gives about \IssueSlopeBOne{} gas per asset under B1 and \IssueSlopeBTwo{} under B2, against a constant \GasIssueOneBZero{} under B0, and an issuance allocating to 32 assets cost \GasIssueThirtyTwoBTwo{} gas under B2. Under B2F each first allocation also records a candidate until the 32-entry list is full; in the benchmark the list filled before the 32-asset tranche, so that tranche cost little more than under B2 and the B2F curve is not linear. When the B2F witness failed, transfer gas rose by about \ScanSlope{} gas per position of the first eligible candidate (\cref{fig:costs}, right), from \ScanFirst{} gas at position \ScanFirstPos{} to \ScanLast{} gas at position \ScanLastPos, about five times the \ScanDirect{} gas of a transfer with a valid witness in the same benchmark.

Renewing an existing claim cost \GasRenewRights{} gas (rights), \GasRenewValuation{} (valuation), \GasRenewClassification{} (classification) and \GasRenewStatus{} (Shariah status). Renewing only the four claims the transfer check reads, the witness's three claims and the Shariah status, kept transfers live in test T17 and cost \GasLiveness{} gas per 30-day cycle, or \GasLivenessYear{} gas over the \CyclesPerYear{} cycles of a year; keeping all 32 assets of a large portfolio current would cost \GasFullYearThirtyTwo{} gas a year (Table~S14 converts these figures into fees). Deploying an evidence configuration cost about \DeployOverheadTwo{} times as much gas as B0. Median transfer times on the simulated network were \LatencyTransferMedMin--\LatencyTransferMedMax{} ms; they describe the simulator on one machine and say nothing about block production or propagation on a public network.

%% file: tables/decision_compact.tex
B0 & 10/23 & 13 & 0 & & 3/4 & 1 & 0 \\
B1 & 20/26 & 6 & 0 & & 3/4 & 0 & 1 \\
B2 & 26/26 & 0 & 0 & & 2/4 & 0 & 2 \\
B2F & 26/26 & 0 & 0 & & 4/4 & 0 & 0 \\

%% file: tables/oracle_supp.tex
G01 & transfer & Witness asset valued at zero while another non-debt asset remains eligible; witness not re-nominated & A & A & \textbf{D}$^{\ast}$ & \textbf{D}$^{\ast}$ & A & \code{NON\_DEBT\_COMPONENT\_UNVERIFIED} \\
G02 & transfer & Witness asset reclassified as debt while another non-debt asset remains eligible; witness not re-nominated & A & A & A & \textbf{D}$^{\ast}$ & A & \code{NON\_DEBT\_COMPONENT\_UNVERIFIED} \\
G03 & transfer & Attested values of all three assets fall to 1\% of their issuance values; the witness remains eligible & A & A & A & A & A & \code{OK} \\
G04 & issue & A new tranche allocates to an asset whose valuation was reduced below the amount already allocated to it & D & \textbf{A}$^{\ast}$ & D & D & D & \code{ALLOCATION\_INVALID} \\

%% file: tables/gas.tex
Deployment, all contracts & 2{,}500{,}415 & 7{,}587{,}314 & 7{,}587{,}362 & 7{,}671{,}778 & +203.4\% & +203.4\% \\
Deployment setup transactions$^{a}$ & 242{,}119 & 598{,}436 & 669{,}518 & 669{,}496 & +147.2\% & +176.5\% \\
First issuance (3 assets, 2 holders) & 140{,}346 & 429{,}197 & 477{,}390 & 567{,}242 & +205.8\% & +240.2\% \\
Issuance, 1 fresh asset, 1 holder & 59{,}772 & 240{,}750 & 279{,}172 & 306{,}608 & +302.8\% & +367.1\% \\
Issuance, 32 fresh assets, 1 holder & 59{,}784 & 1{,}458{,}901 & 1{,}784{,}376 & 1{,}796{,}426 & +2340.3\% & +2884.7\% \\
Transfer, mean of 1{,}000$^{b}$ & 50{,}492 & 78{,}234 & 86{,}326 & 86{,}439 & +54.9\% & +71.0\% \\
Transfer creating a new balance & 67{,}598 & 95{,}340 & 103{,}432 & 103{,}545 & +41.0\% & +53.0\% \\
\midrule
First rights claim for an asset & -- & 97{,}307 & 97{,}307 & 97{,}307 & & \\
First valuation for an asset & -- & 114{,}042 & 114{,}030 & 114{,}042 & & \\
First classification for an asset & -- & -- & 92{,}027 & 92{,}027 & & \\
Valuation renewal (mean of 1{,}000) & -- & 62{,}790 & 62{,}790 & 62{,}790 & & \\
Rights renewal (mean of 100) & -- & 63{,}103 & 63{,}103 & 63{,}104 & & \\
Classification renewal (mean of 100) & -- & -- & 57{,}836 & 57{,}836 & & \\
Shariah status renewal (mean of 100) & -- & -- & 58{,}122 & 58{,}124 & & \\
Suspend (one governor) & -- & 77{,}984 & 77{,}984 & 77{,}984 & & \\
Approve resumption (first vote) & -- & 74{,}835 & 74{,}835 & 74{,}813 & & \\
Approve resumption (quorum vote) & -- & 62{,}063 & 62{,}063 & 62{,}041 & & \\
Nominate witness asset & -- & 51{,}105 & 56{,}235 & 56{,}300 & & \\

%% file: tables/fig_issuance.tex
\addplot coordinates {(1,59.8) (2,59.8) (4,59.8) (8,59.8) (16,59.8) (32,59.8)};
\addplot coordinates {(1,240.8) (2,280.1) (4,358.7) (8,515.8) (16,830.2) (32,1458.9)};
\addplot coordinates {(1,279.2) (2,327.8) (4,424.8) (8,619.1) (16,1007.5) (32,1784.4)};
\addplot coordinates {(1,306.6) (2,377.8) (4,520.2) (8,805.1) (16,1330.1) (32,1796.4)};

%% file: tables/fig_scan.tex
\addplot coordinates {(3,109.1) (4,121.1) (5,133.2) (6,145.2) (7,157.3) (8,169.3) (9,181.4) (10,193.4) (11,205.5) (12,217.5) (13,229.6) (14,241.6) (15,253.7) (16,265.7) (17,277.8) (18,289.8) (19,301.9) (20,313.9) (21,326.0) (22,338.0) (23,350.1) (24,362.1) (25,374.2) (26,386.2) (27,398.3) (28,410.3) (29,422.4) (30,434.4) (31,446.5) (32,458.5)};

%% file: sections/discussion.tex
\section{Discussion}
\label{sec:discussion}

This section answers the four research questions in turn and then states how far the results generalise and where they stop.

\subsection{RQ1: integrating signed asset evidence}
\label{sec:rq1}

Signed claims can be integrated by verifying each one once, when it is submitted, and storing the latest accepted claim per asset and claim type in a registry that the policy engine reads at decision time. Currency is evaluated at each use, so expiry, revocation of a claim or of its signer, and a newer sequence number take effect without any action on the token. The comparison of B0 and B1 isolates the effect of this layer. Of the \OracleBZeroFalseAccept{} primary operations that B0 wrongly allowed, B1 denied all but \OracleBOneFalseAccept, and every remaining false accept depends on Shariah evidence (\cref{sec:results}). Forged or misattributed signatures, replayed or foreign tranches, expired evidence, over-allocation and a compromised key became on-chain denials with explicit reasons, which an identity-based token cannot express. Verifying four signed claims inside each transfer was the alternative; it would have added \GasInlineDelta{} gas to the check, and an inline bundle cannot reveal a later revocation, a newer claim or an encumbrance notice without consulting the same stored state (\cref{sec:evidence,sec:cost}).

The second half of RQ1 has a negative answer that the tests make concrete. Signed claims establish authenticity, currency and eligibility under the policy (\cref{tab:layers}), and nothing more. A valuer's false statement passes every check. The R1 check rests on an allocation co-signed by the wakeel and the trustee, which is a statement about the use of proceeds and not an observed investment; linking allocations to executed acquisitions, for example through payment records signed by a separate role, was not attempted. The claims also say nothing about sufficiency or uniqueness: transfers continued after every attested value fell to 1\% of its issuance value (G03, T16), and a second instrument allocated against an asset the first had already allocated in full (T15). Investors may read ``asset-backed'' as implying coverage; a product that intends continuing coverage would need a separate rule, such as the coverage ratio of RWA-PoB \citep{mafrur2026pob}, with its own evidence.

\subsection{RQ2: encoding one tradability rule}
\label{sec:rq2}

The SAC resolution could be encoded as a versioned policy in which every check is traced to an element of the resolution, an interpretation by the authors or a control the authors added (\cref{tab:trace}). Keeping the issuance test apart from the continuing condition was the central design decision. S05 shows why: after issuance the non-debt share fell to about 6\% while transfers continued, which the resolution permits and a permanent 33\% floor would have blocked. The boundary scenarios fixed the reading of ``at least 33\%'' at 33.00\% (S02, S03). On this encoding B2 agreed with the policy expectation in all \OracleScenarios{} primary scenarios, and in the model-based runs its outcomes, including every transfer and issuance reason code, equalled at every step those of a reference model written separately from the specification.

Two qualifications bound this answer. First, R4 could not be encoded. A token transfer carries no price, so the price consequence of losing the non-debt component is neither observed nor enforced; the prototype substitutes a fail-closed suspension of ordinary transfers, which is the authors' control and not a requirement of the resolution, and S06 and S23 test only that substitute. Second, conformance is measured against the authors' own reading. The scenarios, expectations, reference model and contracts were written by the same authors, so agreement can expose implementation errors but not a misreading of the resolution, and the encoding has not been reviewed by a qualified Shariah adviser.

\subsection{RQ3: behaviour when evidence and composition change}
\label{sec:rq3}

When evidence lapses or composition changes, the controls fail closed, and they do so even when the rule is still satisfied. Expiry, revocation, a contradictory encumbrance notice and a compromised key each stopped the operations that depended on the affected claims, and transfers resumed only after re-attestation and, where needed, re-nomination of the witness (S18, S19). In the adversarial profile, with short validity windows and frequent revocations, encumbrances and suspensions, most operations were denied: under B2 only \ModelTransferOKPctBTwoAdversarial\% of transfers succeeded and \ModelIssuanceDrawnOKBTwoAdversarial{} of \ModelIssuanceDrawnTotalBTwoAdversarial{} randomly drawn issuances were authorised.

The controls also denied operations the rule permits. With a single witness, B2 checks one asset; when that asset lapsed while another non-debt asset still qualified, B2 denied transfers that the resolution permits (G01, G02). The primary scenarios missed this because they re-nominate the witness as a relayer would, and the model-based runs show that the case is not rare: under the same seeds and step mix B2F reduced the denials for an unverified non-debt component from \ModelNonDebtBTwoMaintained{} to \ModelNonDebtBTwoFMaintained{} in the maintained profile. The correction is bounded. B2F scans only the first 32 allocated assets, and T21 shows that an eligible 33rd asset does not keep transfers live. The choice among a prompt relayer, a bounded scan and alternatives such as several nominated witnesses depends on portfolio size and on the cost of a halted market; the prototype measures only the first two.

Some changes leave the controls indifferent by design. A fall in attested values does not stop transfers (G03), although it limits new allocations to the devalued asset (G04), and the agent's forced transfer moves units while the continuing-component check is failing (T18).

\subsection{RQ4: gas, attestation and governance costs}
\label{sec:rq4}

The evidence layer adds a constant cost to each transfer and a cost per asset to each issuance. Relative to B0, B2 added \GasTransferDeltaTwo{} gas (\OverheadTransferTwo\%) to a transfer whatever the size of the portfolio, which under the assumed gas prices amounts to between \CostTransferDeltaLow{} and \CostTransferDeltaHigh{} units of the native token; whether that matters depends on the chain, the token price and trading frequency, none of which the study observes. Under B2, issuance cost grew by about \IssueSlopeBTwo{} gas per freshly evidenced asset, and deploying an evidence configuration cost about \DeployOverheadTwo{} times as much gas as B0. B2F's liveness is paid for at transfer time, at about \ScanSlope{} gas per scanned position.

Attestation cost recurs with time. Keeping every asset eligible requires $3n+1$ claim renewals per validity period for $n$ assets, whereas the four renewals in T17 kept transfers live at \GasLiveness{} gas per 30-day cycle; full renewal matters for issuance capacity, for B2F's candidates and for the accuracy of the record. The validity window trades this cost against the period during which a sold or impaired asset keeps counting, and the 30-day maximum is a parameter of the prototype, not a recommendation.

The governance cost is borne in market continuity and in concentrated authority. Every fail-closed check converts an attestor's delay, a governor's suspension or a relayer's absence into a halt of ordinary transfers that holders did not cause. One registry governor key controls attestor onboarding, any one governor can suspend trading, and the agent can force transfers around the policy; logging makes misuse attributable after the fact without preventing it. Placing onboarding behind the policy quorum and timelock, separating the agent, governor and administrator keys, requiring a quorum for forced transfers and letting unconfirmed suspensions lapse would reduce reliance on single keys at the price of slower responses to emergencies. None of these measures is implemented.

\subsection{Generalisation and limitations}
\label{sec:generalisation}

The evidence registry is not specific to sukuk: rights, valuation and encumbrance claims apply to other asset-backed tokens. Compared with RWA-PoB, the design replaces one aggregate snapshot signed by several roles with per-asset claims, each signed by the role competent for it, and adds a sukuk-specific rule with separate issuance and continuing conditions. The policy engine is specific to the SAC resolution. The AAOIFI position would need composition tracking over time and a restoration timer, an ijarah sukuk would need lease-status claims, and adapting the engine as an ERC-3643 compliance module would cover transfers but not issuance, since ERC-3643 minting bypasses compliance; none of these has been implemented.

The results carry six limitations. The encoding is the authors' reading of one resolution and has not been reviewed by a qualified Shariah scholar or a Malaysian securities lawyer. The scenarios, expectations, reference model and contracts were written by the same authors, so agreement demonstrates conformance to the specification and not its correctness. The oracle is a constructed set, and its results are not error rates on real instruments. The instrument and evidence are synthetic and B0 is a simplified baseline. Gas and timing come from a local simulated chain and fees from assumed prices; no public-network deployment was made. Finally, the contracts have not been audited or formally verified, and claims, signers and allocations would be public on a public chain, since the claim format does not support selective disclosure \citep{flamini2024cryptographic}.

%% file: sections/conclusion.tex
\section{Conclusion}
\label{sec:conclusion}

This study built and evaluated smart contracts that make the issuance and ordinary transfer of a tokenised Sukuk Wakalah depend on current, role-attributed evidence about individual assets and on an explicit encoding of the SAC 2023 resolution. In the \OracleScenarios{} primary scenarios constructed for the purpose, the full policy agreed with the specified oracle in all \OracleScenarios, while the simplified ERC-3643-style baseline and the evidence layer alone accepted operations the policy denies. The 33\% minimum was tested only at issuance and the non-debt component only at transfer, as the resolution distinguishes them. The single-witness design falsely rejected transfers when the witness lapsed while another non-debt asset remained eligible; a fallback scan over the first 32 allocated assets removed these rejections in the tested cases at a cost that grows with the scan position, and left them in place beyond the cap. The policy check added \GasTransferDeltaTwo{} gas to an ordinary transfer, and renewing the four claims that the transfer check reads kept transfers live.

The contracts establish who attested what and whether the attestation is current, not that the attested facts are true, that proceeds were invested or that the portfolio covers the certificates outstanding, and they cannot observe the prices at which certificates trade. The encoding, including the decision to suspend transfers when the non-debt component is not evidenced, requires review by qualified Shariah and legal experts before it can be relied on. A settlement mechanism that observes trade prices, links between allocations and executed payments, and a policy module for the AAOIFI continuing floor are the natural next steps.

%% file: sections/declarations.tex
\section*{Data and code availability}

The Solidity contracts, TypeScript and Solidity tests, synthetic fixtures, scenario oracle, coverage matrix, raw outputs, static-analysis and mutation reports, environment record and the scripts that generate every table and figure in this paper and its supplementary material are provided as a supplementary code archive with this submission, under the MIT licence. Its README gives the commands that reproduce each reported result, including the full model-based runs. All data are synthetic.

\section*{Declaration of generative AI and AI-assisted technologies in the writing process}

The authors acknowledge the use of OpenAI's ChatGPT
(GPT-6, October 2026) for grammar checking and language
refinement during manuscript preparation. All scientific
content, analysis, interpretations, and conclusions
remain the sole responsibility of the authors.

%% file: sections/abbreviations.tex
\section{Abbreviations and identifiers}
\label{app:abbrev}

\begin{table}[!h]
\centering
\caption{Abbreviations, configuration names and identifiers used in the paper.}
\label{tab:abbrev}
\small
\begin{tabularx}{\textwidth}{@{}lX@{}}
\toprule
Term & Meaning \\
\midrule
\multicolumn{2}{@{}l}{\textit{Organisations and standards}} \\
AAOIFI & Accounting and Auditing Organization for Islamic Financial Institutions \\
EIP & Ethereum Improvement Proposal; EIP-712 defines typed, domain-separated signatures \\
ERC & Ethereum Request for Comments; ERC-20 defines fungible tokens, ERC-1271 signature validation by contract wallets, ERC-3643 permissioned security tokens \\
IFSB & Islamic Financial Services Board \\
RWA-PoB & Real-world asset proof of backing, a framework for tokenised Treasury products \citep{mafrur2026pob} \\
SAC & Shariah Advisory Council of the Securities Commission Malaysia \\
W3C & World Wide Web Consortium \\
\addlinespace
\multicolumn{2}{@{}l}{\textit{Configurations}} \\
B0 & Simplified baseline in the style of ERC-3643 that checks investor identity only \\
B1 & B0 plus signed asset and allocation evidence, without Shariah rules \\
B2 & B1 plus the encoded SAC policy, with the continuing condition checked through one witness asset \\
B2F & B2 plus a fallback scan over up to 32 allocated assets when the witness fails \\
\addlinespace
\multicolumn{2}{@{}l}{\textit{Identifiers}} \\
R1--R4 & Elements of the SAC resolution (\cref{sec:policy}) \\
I1--I8 & Safety invariants (\cref{sec:trust}) \\
RQ1--RQ4 & Research questions (\cref{sec:intro}) \\
S01--S26 & Primary oracle scenarios (\cref{sec:oracle}) \\
G01--G04 & Supplementary scenarios (\cref{sec:oracle}, \cref{tab:supp}) \\
T01--T21 & Targeted tests (\cref{sec:assurance_design}) \\
FA, FR & False accept, false reject (\cref{sec:oracle}) \\
\addlinespace
\multicolumn{2}{@{}l}{\textit{Other}} \\
EDR & Ethereum Development Runtime, the simulated network of Hardhat \\
gwei & $10^{-9}$ of the native token, the unit of gas prices \\
\bottomrule
\end{tabularx}
\end{table}
\FloatBarrier

%% file: supplement/body.tex

\section{Related work}
\label{sec:s-priorart}

\Cref{tab:priorart} compares the closest works identified in Section~2 with the present design. The comparison is limited to the studies cited there and does not result from a systematic search.

\input{supplement/priorart}
\FloatBarrier

\section{Policy, invariants, parties and threats}
\label{sec:s-policy}

\Cref{tab:invariants} states the eight safety invariants tested, where each is enforced and its scope; none has been proved formally. \Cref{tab:roles} lists the parties and their on-chain authority, and \cref{tab:threats} the threats TH1--TH11 with their mitigations and the scenarios and tests that exercise them.

\input{supplement/invariants}
\input{supplement/roles}
\input{supplement/threats}
\FloatBarrier

\section{Scenario oracle and coverage}
\label{sec:s-oracle}

\Cref{tab:oracle} gives every primary scenario and every configuration's decision. \Cref{tab:coverage_elements} shows which scenarios exercise each resolution element, invariant and threat, and \cref{tab:coverage_matrix} gives the full mapping. The R4 price treatment is not tested by any scenario, because a token transfer carries no price; S06 (blocked) and S23 (allowed) exercise only the authors' transfer control (Section~3.1). The four supplementary scenarios are in Table~5 of the manuscript.

\input{supplement/oracle}
\input{supplement/coverage_elements}
\input{supplement/coverage_matrix}
\FloatBarrier

\section{Targeted tests}
\label{sec:s-tests}

\input{supplement/tests}
\FloatBarrier

\section{Model-based differential testing}
\label{sec:s-mb}

\Cref{tab:mb} gives the distribution of outcomes and \cref{tab:mb_other} the other steps drawn. A shorter run of five seeds of 250 steps in the adversarial profile, kept as a quick check, gave under B2 \RepTransferOKBTwo{} successful transfers of \RepTransferTotalBTwo{} and \RepIssueDrawnOKBTwo{} authorised issuances of \RepIssueDrawnTotalBTwo{} randomly drawn, with contract and model in agreement at every step.

\input{supplement/mb}
\input{supplement/mb_other}
\FloatBarrier

\section{Static analysis, mutation checks and line coverage}
\label{sec:s-static}

\input{supplement/static}
\input{supplement/mutations}
\input{supplement/linecoverage}
\FloatBarrier

\section{Costs, contract sizes and latency}
\label{sec:s-costs}

\Cref{tab:costs} converts selected gas figures of Table~6 into fees in native-token units under hypothetical gas prices. \Cref{tab:contracts,tab:bytecode} list the contracts and their sizes, and \cref{tab:latency} the wall-clock latency on the simulated network.

\input{supplement/costs}
\input{supplement/contracts}
\input{supplement/bytecode}
\input{supplement/latency}
\FloatBarrier

\section{Claim schemas and lifecycle}
\label{sec:s-schemas}

Both signed structures use an EIP-712 domain with name \code{TokenizedSukukEvidence}, version \code{1}, the chain identifier and, as verifying contract, the address of the evidence registry (\cref{lst:types}). \Cref{tab:claimtypes} lists the claim types and \cref{fig:lifecycle} the instrument lifecycle.

\input{supplement/types}
\input{supplement/claimtypes}
\input{supplement/lifecycle}
\FloatBarrier

\section{Reproduction}
\label{sec:s-reproduce}

The supplementary code archive was run with Node.js 22 and pins its npm dependencies through \texttt{package-lock.json} (Hardhat \EnvHardhat, viem \EnvViem, OpenZeppelin Contracts \EnvOZ, solc \EnvSolc). From its root, \texttt{npm ci} followed by \texttt{scripts/reproduce.sh} runs, in order, the instrumented coverage run; the measured run of all suites, including the oracle, the model-based runs at \ModelSeeds{} seeds $\times$ \ModelSteps{} steps in both profiles for B1, B2 and B2F, the fuzz properties and the gas benchmark; the five-seed replication; Slither and the mutation checks; and the generation of every table and of this document. The README lists the individual commands and their outputs. The Slither findings before the fixes and the mutation results before tests T19 and T20 were added are archived outputs of the earlier code and cannot be regenerated from the current sources.

%% file: supplement/priorart.tex
\begin{table}[htbp]
\centering
\caption{Comparison with the closest prior work, as reported in each source. ``--'' means not reported in the source.}
\label{tab:priorart}
\small
\begin{tabularx}{\textwidth}{@{}P{2.5cm}P{2.3cm}LLP{2.1cm}P{2.2cm}@{}}
\toprule
Work & Artefact & Asset or rights evidence & Signer roles and claim currency & Sukuk rule encoded & Evaluation \\
\midrule
\citet{khan2022tokenization} & Solidity prototype, basic issuance & -- & -- & -- & gas per function; cost comparison \\
\citet{shaikh2019blockchained} & conceptual design & traceability of financing to assets (proposed) & parties named; no key model & -- & none \\
\citet{moppratama2024property} & stakeholder and data model & property, financial and compliance records & stakeholder roles & -- & practitioner interviews \\
ERC-3643 \citep{lebrun2021erc3643} & standard and reference code & investor identity claims only & trusted claim issuers per topic & -- (generic rules) & -- \\
RWA-PoB \citep{mafrur2026pob} & Solidity prototype & one aggregate reserve and liability snapshot & five roles sign the same snapshot; freshness and epochs & -- (Treasury products) & scenario, invariant, adversarial and gas tests \\
This paper & Solidity prototype & per-asset rights, valuation, encumbrance and classification claims; tranche allocations & six roles, one per key, role bound to claim type; validity, sequence, revocation, domain binding & SAC 2023 Wakalah: issuance minimum and continuing component & scenario oracle over four configurations; model-based, fuzz, static and mutation checks; gas \\
\bottomrule
\end{tabularx}
\end{table}

%% file: supplement/invariants.tex
\begin{table}[htbp]
\centering
\caption{Safety invariants, enforcement points and scope. Tests: identifiers of \cref{tab:tests}; MB, checked after every step of the model-based runs; F, Solidity fuzz property.}
\label{tab:invariants}
\footnotesize
\begin{tabularx}{\textwidth}{@{}p{0.55cm}LP{3.2cm}P{2.6cm}@{}}
\toprule
ID & Invariant and scope & Enforced in & Tests \\
\midrule
I1 & No issuance when the issuance minimum fails or required evidence is invalid. & \code{authorizeIssuance} & T01, T02, T05, MB, F \\
I2 & No ordinary transfer when the non-debt component is not supported by current evidence for the witness (B2) or any recorded candidate (B2F). & \code{canTransfer} & T03, T04, T06, T08, T21, MB \\
I3 & No issuance, policy change or attestor change by an unauthorised key; approvals are logged. & roles, quorum, timelock & T05, T10 \\
I4 & Revoked, expired, older, wrong-chain, wrong-contract or wrong-instrument claims cannot update authoritative state. & \code{submitAttestation}; currency at use & T06, T07, T11, T12, F \\
I5 & Within one instrument, an asset cannot be allocated beyond its current attested value or counted twice in a tranche; allocations do not exceed proceeds. Pledges in other instruments are not visible. & \code{\_checkAllocation} & T08, T15, T19, T20, MB, F \\
I6 & Issued units equal supply plus pending plus settled redemptions; no redemption or settlement beyond the outstanding amount. & \code{requestRedemption}, \code{confirmSettlement} & T13, MB, F \\
I7 & Every balance change passes the identity and policy pre-check, except three logged privileged paths: authorised issuance, redemption burns and the agent's forced transfer. & \code{\_update} hook & T09, T18, MB \\
I8 & Suspension and forced transfers are attributable; policy changes do not rewrite recorded tranches. & events, version records & T10, T18 \\
\bottomrule
\end{tabularx}
\end{table}

%% file: supplement/roles.tex
\begin{table}[htbp]
\centering
\caption{Parties, their on-chain authority and what the system trusts them for.}
\label{tab:roles}
\footnotesize
\begin{tabularx}{\textwidth}{@{}P{2.6cm}LL@{}}
\toprule
Party & On-chain authority & Trusted for \\
\midrule
Governors (3 keys, quorum 2) & propose and approve policy versions, and activate them no earlier than one day after the proposal; suspend (any one); resume (quorum); onboard and revoke attestors and revoke any stored claim (one registry governor key in the prototype, which as registry administrator can also grant the governor role) & honest onboarding; no collusion \\
Agent (issuer operator) & submit issuance with tranche evidence; pause; freeze; forced transfer with a justification hash; manage identity entries & investor eligibility data; legitimate use of forced transfers \\
Wakeel and trustee & co-sign each tranche allocation & truthful statement of the allocation of proceeds; independent check \\
Legal verifier & rights claims & existence and scope of the portfolio's rights \\
Valuer & valuation claims & fair valuation \\
Custodian & encumbrance notices & timely notice of liens \\
Shariah adviser & asset classification; instrument Shariah status & Shariah judgements \\
Settlement role & confirm redemption payments & that payment occurred \\
Relayer (anyone) & submit signed claims; nominate a witness & nothing; it cannot alter signed content \\
\bottomrule
\end{tabularx}
\end{table}

%% file: supplement/threats.tex
\begin{table}[htbp]
\centering
\caption{Threats, mitigations in the prototype and the scenarios and tests that exercise them.}
\label{tab:threats}
\footnotesize
\begin{tabularx}{\textwidth}{@{}p{0.75cm}LLP{2.4cm}@{}}
\toprule
ID & Threat & Mitigation & Evidence \\
\midrule
TH1 & Ineligible investor or operator bypasses transfer rules, including via \code{transferFrom} & every balance change passes the pre-check except three logged privileged paths & S16, T09, I7 tests, MB \\
TH2 & Forged signature, unregistered key, key outside its role, altered payload & one role per key; role bound to claim type; EIP-712 digest over all fields; EOA and ERC-1271 verification & S07, S08, T05 \\
TH3 & Replay of a claim or tranche; cross-instrument, cross-chain or cross-contract replay & per-subject sequence numbers; tranche identifiers used once; instrument in payload; chain and contract in domain & S12, S13, T07 \\
TH4 & Stale or withheld evidence & maximum validity of 30 days; currency evaluated at use; fail-closed transfers & S09, S10, T06 \\
TH5 & Compromised attestor key & governor revokes the attestor; all its claims except encumbrance notices stop counting; new key and fresh claims (unless the key has already submitted the largest sequence number for a claim type) & S11, S18, S19, T11 \\
TH6 & Out-of-order updates & sequence numbers decide which claim is authoritative & S20, T12 \\
TH7 & Contradictory rights and encumbrance evidence & conflict event; encumbered assets excluded; notices do not lapse & S14, T08 \\
TH8 & Double counting of assets within the instrument & cumulative allocation bounded by current attested value; sorted, duplicate-free allocations; total within proceeds & S15, G04, T08, T19, T20, MB \\
TH9 & Silent or unilateral policy change & versioned policy; quorum; timelock; no retroactive effect; version-bound Shariah attestations & S17, T10 \\
TH10 & Double redemption or unaccounted obligations & burn to pending; settlement bounded by pending amount & S21, S22, T13, MB \\
TH11 & Emergency (dispute, suspected compromise) & any governor suspends; quorum resumes; votes scoped to the suspension epoch & S25, lifecycle tests \\
\midrule
\multicolumn{4}{@{}p{\textwidth}@{}}{\emph{Not mitigated:} false content under a valid signature; outdated input within the validity window; collusion among roles; pledges of the same asset in other instruments (T15); decline in aggregate coverage (G03, T16); the agent's forced transfers (T18); concentration of governor keys in the test deployment.} \\
\bottomrule
\end{tabularx}
\end{table}

%% file: supplement/oracle.tex
\begin{table}[!htbp]
\centering
\caption{Primary scenario oracle (S01--S26). Policy: outcome required by the authors' policy encoding. A, allowed; D, denied; --, not applicable. $^{\ast}$ marks a decision that differs from the policy. The last column gives B2's reason code or error.}
\label{tab:oracle}
\scriptsize
\setlength{\tabcolsep}{2.6pt}
\begin{tabularx}{\textwidth}{@{}p{0.5cm}p{1.15cm}Lcccccp{3.55cm}@{}}
\toprule
ID & Operation & Scenario & Policy & B0 & B1 & B2 & B2F & B2 reason \\
\midrule
\input{tables/oracle.tex} 
\bottomrule
\end{tabularx}
\end{table}

%% file: tables/oracle.tex
S01 & issue & Valid credentials; 35\% of proceeds allocated to non-debt assets & A & A & A & A & A & \code{OK} \\
S02 & issue & Valid credentials; exactly 33.00\% non-debt (boundary) & A & A & A & A & A & \code{OK} \\
S03 & issue & Valid credentials; 32.99\% non-debt (boundary) & D & \textbf{A}$^{\ast}$ & \textbf{A}$^{\ast}$ & D & D & \code{NON\_DEBT\_RATIO\_BELOW\_MINIMUM} \\
S04 & issue & One allocated asset reclassified as debt by the Shariah adviser so the non-debt share falls to 25\% & D & \textbf{A}$^{\ast}$ & \textbf{A}$^{\ast}$ & D & D & \code{NON\_DEBT\_RATIO\_BELOW\_MINIMUM} \\
S05 & transfer & After issuance the non-debt share falls to about 6\% but a non-debt asset remains & A & A & A & A & A & \code{OK} \\
S06 & transfer & After issuance every non-debt asset is disposed (value attested as zero) & D & \textbf{A}$^{\ast}$ & \textbf{A}$^{\ast}$ & D & D & \code{NON\_DEBT\_COMPONENT\_UNVERIFIED} \\
S07 & issue & Tranche allocation co-signed by an unregistered key instead of the trustee & D & \textbf{A}$^{\ast}$ & D & D & D & \code{TRANCHE\_INVALID} \\
S08 & issue & Allocation exceeds attested value after a forged valuation claim was rejected & D & \textbf{A}$^{\ast}$ & D & D & D & \code{ALLOCATION\_INVALID} \\
S09 & transfer & All evidence has expired (time advanced past validity) & D & \textbf{A}$^{\ast}$ & D & D & D & \code{SHARIAH\_STATUS\_INVALID} \\
S10 & transfer & Transfer at the exact expiry second of the evidence (inclusive bound) & A & A & A & A & A & \code{OK} \\
S11 & transfer & Rights credentials of every non-debt asset revoked by the legal verifier & D & \textbf{A}$^{\ast}$ & \textbf{A}$^{\ast}$ & D & D & \code{NON\_DEBT\_COMPONENT\_UNVERIFIED} \\
S12 & issue & Second issuance replaying an already used tranche allocation & D & \textbf{A}$^{\ast}$ & D & D & D & \code{TRANCHE\_REPLAYED} \\
S13 & issue & Tranche allocation signed for another chain id & D & \textbf{A}$^{\ast}$ & D & D & D & \code{TRANCHE\_INVALID} \\
S14 & transfer & Custodian notices of encumbrance contradict the legal verifier on both non-debt assets & D & \textbf{A}$^{\ast}$ & \textbf{A}$^{\ast}$ & D & D & \code{NON\_DEBT\_COMPONENT\_UNVERIFIED} \\
S15 & issue & Second tranche re-uses value already allocated to an asset & D & \textbf{A}$^{\ast}$ & D & D & D & \code{ALLOCATION\_INVALID} \\
S16 & transfer & Transfer to a wallet that is not in the identity registry & D & D & D & D & D & \code{RECEIVER\_NOT\_VERIFIED} \\
S17 & governance & Policy activation with one governor approval (quorum is two) & D & -- & D & D & D & \code{QuorumNotReached} \\
S18 & transfer & Legal verifier key revoked after compromise; no remediation yet & D & \textbf{A}$^{\ast}$ & D & D & D & \code{NON\_DEBT\_COMPONENT\_UNVERIFIED} \\
S19 & transfer & After remediation: new legal key re-attests and witness re-nominated & A & A & A & A & A & \code{OK} \\
S20 & evidence & Older valuation (lower sequence) submitted after a newer one & D & -- & D & D & D & \code{StaleOrReplayedClaim} \\
S21 & redeem & After maturity a holder requests redemption beyond remaining balance & D & D & D & D & D & \code{ERC20InsufficientBalance} \\
S22 & settle & Settlement confirmation exceeding the pending claim & D & D & D & D & D & \code{InvalidRedemption} \\
S23 & transfer & Ordinary transfer between verified holders with current evidence & A & A & A & A & A & \code{OK} \\
S24 & issue & Issuance while the Shariah status attestation says not endorsed & D & \textbf{A}$^{\ast}$ & \textbf{A}$^{\ast}$ & D & D & \code{SHARIAH\_STATUS\_INVALID} \\
S25 & transfer & Instrument suspended by one governor (dispute) & D & -- & D & D & D & \code{INSTRUMENT\_SUSPENDED} \\
S26 & transfer & Ordinary transfer after maturity & D & D & D & D & D & \code{INSTRUMENT\_MATURED} \\

%% file: supplement/coverage_elements.tex
\begin{table}[htbp]
\centering
\caption{Coverage of resolution elements (R), invariants (I) and threats (TH, \cref{tab:threats}) by the scenario oracle. No scenario tests the R4 price treatment; S06 (transfers blocked once the non-debt component is lost) and S23 (an ordinary transfer allowed while R1 and R3 hold) exercise only the authors' transfer control, which serves as a proxy.}
\label{tab:coverage_elements}
\footnotesize
\begin{tabularx}{\textwidth}{@{}p{1.0cm}p{0.6cm}LP{2.6cm}@{}}
\toprule
Element & $n$ & Primary scenarios & Supplementary \\
\midrule
\input{tables/coverage_elements.tex} 
\bottomrule
\end{tabularx}
\end{table}

%% file: tables/coverage_elements.tex
R1 & 4 & S01, S02, S03, S04 & -- \\
R2 & 1 & S05 & G03 \\
R3 & 6 & S06, S11, S14, S18, S19, S23 & G01, G02 \\
R4 & 0 & price treatment not testable (no price on-chain); transfer control as proxy only: S06, S23 & -- \\
I1 & 7 & S01, S02, S03, S04, S07, S08, S24 & -- \\
I2 & 7 & S05, S06, S09, S10, S11, S14, S23 & G01, G02, G03 \\
I3 & 2 & S07, S17 & -- \\
I4 & 8 & S09, S10, S11, S12, S13, S18, S19, S20 & -- \\
I5 & 2 & S08, S15 & G04 \\
I6 & 2 & S21, S22 & -- \\
I7 & 2 & S16, S23 & -- \\
I8 & 1 & S25 & -- \\
TH1 & 1 & S16 & -- \\
TH2 & 2 & S07, S08 & -- \\
TH3 & 2 & S12, S13 & -- \\
TH4 & 2 & S09, S10 & -- \\
TH5 & 3 & S11, S18, S19 & -- \\
TH6 & 1 & S20 & -- \\
TH7 & 1 & S14 & -- \\
TH8 & 1 & S15 & G04 \\
TH9 & 1 & S17 & -- \\
TH10 & 2 & S21, S22 & -- \\
TH11 & 1 & S25 & -- \\

%% file: supplement/coverage_matrix.tex
\begin{table}[htbp]
\centering
\caption{Mapping of each oracle scenario to resolution elements, invariants, threats, lifecycle stage and condition. R4 (proxy): the scenario exercises only the authors' transfer control (blocked in S06, allowed in S23), not the R4 price treatment, which is not tested. pos.: positive control; neg.: negative case; bound.: boundary case with the expected policy outcome (A, allow; D, deny).}
\label{tab:coverage_matrix}
\footnotesize
\begin{tabular}{@{}llllll@{}}
\toprule
Scenario & Resolution & Invariants & Threats & Stage & Condition \\
\midrule
\input{tables/coverage_matrix.tex} 
\bottomrule
\end{tabular}
\end{table}

%% file: tables/coverage_matrix.tex
S01 & R1 & I1 & -- & issuance & pos. \\
S02 & R1 & I1 & -- & issuance & bound. (A) \\
S03 & R1 & I1 & -- & issuance & bound. (D) \\
S04 & R1 & I1 & -- & issuance & neg. \\
S05 & R2 & I2 & -- & transfer & pos. \\
S06 & R3, R4 (proxy) & I2 & -- & transfer & neg. \\
S07 & -- & I1, I3 & TH2 & issuance & neg. \\
S08 & -- & I1, I5 & TH2 & issuance & neg. \\
S09 & -- & I2, I4 & TH4 & transfer & neg. \\
S10 & -- & I2, I4 & TH4 & transfer & bound. (A) \\
S11 & R3 & I2, I4 & TH5 & transfer & neg. \\
S12 & -- & I4 & TH3 & issuance & neg. \\
S13 & -- & I4 & TH3 & issuance & neg. \\
S14 & R3 & I2 & TH7 & transfer & neg. \\
S15 & -- & I5 & TH8 & issuance & neg. \\
S16 & -- & I7 & TH1 & transfer & neg. \\
S17 & -- & I3 & TH9 & governance & neg. \\
S18 & R3 & I4 & TH5 & transfer & neg. \\
S19 & R3 & I4 & TH5 & transfer & pos. \\
S20 & -- & I4 & TH6 & evidence & neg. \\
S21 & -- & I6 & TH10 & redemption & neg. \\
S22 & -- & I6 & TH10 & redemption & neg. \\
S23 & R3, R4 (proxy) & I2, I7 & -- & transfer & pos. \\
S24 & -- & I1 & -- & issuance & neg. \\
S25 & -- & I8 & TH11 & transfer & neg. \\
S26 & -- & -- & -- & transfer & neg. \\
G01 & R3 & I2 & -- & transfer & pos. \\
G02 & R3 & I2 & -- & transfer & pos. \\
G03 & R2 & I2 & -- & transfer & bound. (A) \\
G04 & -- & I5 & TH8 & issuance & neg. \\

%% file: supplement/tests.tex
\begin{table}[htbp]
\centering
\caption{Targeted tests by property. All passed in the reported run. Configurations are those under which the test runs.}
\label{tab:tests}
\footnotesize
\begin{tabularx}{\textwidth}{@{}p{1.3cm}LP{1.9cm}@{}}
\toprule
ID & Property tested & Config. \\
\midrule
T01 & issuance with at least 33\% non-debt allocation is authorised, including exactly 33.00\% & B2 \\
T02 & issuance below 33\% (32.99\%) is denied; an asset reclassified as debt does not count & B2 \\
T03 & share falls below 33\% after issuance with a non-debt asset remaining: transfers allowed; if the witness is invalidated and not re-nominated, transfers fail closed & B2 \\
T03-F & the fallback scan permits transfers without re-nomination while a candidate is eligible, denies when none is, and records at most 32 candidates & B2F \\
T21 & with 33 assets allocated (32 non-debt, one debt) and only the 33rd eligible, transfers are denied (the 33rd is not a candidate); re-nominating it as witness restores transfers & B2, B2F \\
T04 & loss of every non-debt asset denies transfers; a debt asset cannot serve as witness; transfers resume when a new non-debt asset is evidenced & B2 \\
-- & current Shariah status required for issuance and transfer & B2 \\
T05 & unregistered key, wrong role, altered payload, misattributed signature, issuer-side key in a verification role, tranche without wakeel and trustee, malformed claims, invalid attestor registration; ERC-1271 signer accepted & B2 \\
T06 & expired, future-dated and over-long claims and tranche allocations rejected; transfers denied when relied-on evidence expires; revoked rights claim stops counting & B1, B2 \\
T07 & replay and older sequence; other instrument, chain or verifying contract; reused tranche & B2 \\
T08, I5 & conflicting encumbrance flagged and excluded; notice persists after expiry; no allocation beyond attested value or proceeds; unsorted or duplicate allocation rejected; only the bound token can request authorisation & B2 \\
T09 & transfer to a wallet not in the identity registry denied & all \\
I7, I8 & \code{transferFrom} checked; issuance only through the authorised path and to verified wallets; forced transfer needs agent, justification and verified receiver and is logged & B2 \\
-- & compliance contract binds once to a non-zero token; pause, freeze, identity removal and expiry deny transfers & all \\
T10 & policy change without quorum, before the timelock or by a non-governor rejected; activation is prospective and fails closed until re-attestation; malformed or closed proposals and other-version tranches rejected & B2 \\
-- & one governor suspends; resumption needs a quorum; no stacked suspension; votes do not carry over & B2 \\
T11 & attestor compromise: revocation invalidates its claims; key cannot sign or be re-registered; remediation with a new key & B2 \\
T12 & out-of-order updates in one block cannot overwrite a newer claim & B2 \\
T13 & partial redemption and settlement keep $\text{issued} = \text{supply} + \text{pending} + \text{settled}$; no over-redemption or over-settlement & all \\
T15 & a second instrument can allocate against an asset already fully allocated in the first & B2 \\
T16 & with all valuations at 1\%, transfers continue and new allocation to a devalued asset is rejected & B2, B2F \\
T17 & renewing only the witness's three claims and the Shariah status keeps transfers live; allocation to a lapsed asset is rejected & B2 \\
T18 & the agent's forced transfer succeeds while the continuing-component check fails, and emits an attributable event & B2 \\
T19 & allocation exactly equal to the attested value is authorised; one unit more is rejected & B1, B2 \\
T20 & allocations summing to the proceeds are authorised; a larger sum is rejected & B1, B2 \\
\midrule
Oracle & \OracleScenarios{} primary and \OracleSuppScenarios{} supplementary scenarios under four configurations & all \\
MB & seeded model-based runs, two profiles & B1, B2, B2F \\
F & Solidity fuzz properties, \FuzzRuns{} runs each & B2 engine (issuance); B0 token (redemption); registry only (claims) \\
T14 & gas and latency benchmark & all \\
\midrule
 & \textbf{Total: \TestsPassed{} tests (\TestsTS{} TypeScript, \TestsSolidity{} Solidity fuzz), \TestsFailed{} failures} & \\
\bottomrule
\end{tabularx}
\end{table}

%% file: supplement/mb.tex
\begin{table}[htbp]
\centering
\caption{Outcomes in the model-based runs (\ModelSeeds{} seeds $\times$ \ModelSteps{} steps per profile and configuration). Contract and reference model agreed on every outcome. For redemption and settlement the number of submitted steps is shown. Issuance counts include the fixed first tranche of each walk (one per seed), which is always authorised.}
\label{tab:mb}
\scriptsize
\setlength{\tabcolsep}{3.5pt}
\begin{tabular}{@{}llrrrrrr@{}}
\toprule
 & & \multicolumn{3}{c}{Adversarial profile} & \multicolumn{3}{c}{Maintained profile} \\
\cmidrule(lr){3-5}\cmidrule(l){6-8}
Operation & Outcome (reason) & B1 & B2 & B2F & B1 & B2 & B2F \\
\midrule
\input{tables/model_based.tex} 
\bottomrule
\end{tabular}
\end{table}

%% file: tables/model_based.tex
Transfer & \code{OK} & 180 & 78 & 95 & 1388 & 1283 & 1382 \\
 & \code{BACKING\_EVIDENCE\_INVALID} & 384 & -- & -- & 81 & -- & -- \\
 & \code{ERC20InsufficientBalance} & 64 & 49 & 50 & 155 & 172 & 176 \\
 & \code{INSTRUMENT\_MATURED} & 1252 & 777 & 777 & 1269 & 954 & 954 \\
 & \code{INSTRUMENT\_SUSPENDED} & 534 & 776 & 776 & 67 & 263 & 263 \\
 & \code{NON\_DEBT\_COMPONENT\_UNVERIFIED} & -- & 418 & 400 & -- & 222 & 119 \\
 & \code{RECEIVER\_NOT\_VERIFIED} & 394 & 489 & 489 & 571 & 644 & 644 \\
 & \code{SHARIAH\_STATUS\_INVALID} & -- & 210 & 210 & -- & 38 & 38 \\
 & \textit{total} & 2808 & 2797 & 2797 & 3531 & 3576 & 3576 \\
\midrule
Issuance & \code{OK} & 50 & 24 & 24 & 217 & 128 & 128 \\
 & \code{ALLOCATION\_INVALID} & 369 & 310 & 310 & 670 & 663 & 663 \\
 & \code{INSTRUMENT\_SUSPENDED} & 288 & 447 & 447 & 51 & 150 & 150 \\
 & \code{InvalidIssuance} & 530 & 334 & 334 & 544 & 387 & 387 \\
 & \code{NON\_DEBT\_RATIO\_BELOW\_MINIMUM} & -- & 22 & 22 & -- & 122 & 122 \\
 & \code{SHARIAH\_STATUS\_INVALID} & -- & 112 & 112 & -- & 19 & 19 \\
 & \textit{total} & 1237 & 1249 & 1249 & 1482 & 1469 & 1469 \\
\midrule
Redemption & \textit{submitted} & 164 & 117 & 117 & 133 & 98 & 98 \\
\midrule
Settlement & \textit{submitted} & 116 & 72 & 72 & 119 & 92 & 92 \\

%% file: supplement/mb_other.tex
\begin{table}[!htbp]
\centering
\caption{Other steps drawn in the model-based runs (\ModelSeeds{} seeds $\times$ \ModelSteps{} steps per profile and configuration).}
\label{tab:mb_other}
\footnotesize
\setlength{\tabcolsep}{4pt}
\begin{tabular}{@{}lrrrrrr@{}}
\toprule
 & \multicolumn{3}{c}{Adversarial profile} & \multicolumn{3}{c}{Maintained profile} \\
\cmidrule(lr){2-4}\cmidrule(l){5-7}
Step & B1 & B2 & B2F & B1 & B2 & B2F \\
\midrule
\input{tables/model_based_other.tex} 
\bottomrule
\end{tabular}
\end{table}

%% file: tables/model_based_other.tex
valuation updates & 1240 & 1258 & 1258 & 1136 & 1170 & 1170 \\
rights or classification re-attestations & 802 & 810 & 810 & 1047 & 1074 & 1074 \\
custodian notices & 544 & 525 & 525 & 182 & 181 & 181 \\
revocation attempts & 529 & 500 & 500 & 101 & 97 & 97 \\
Shariah status steps (no effect under B1) & 501 & 485 & 485 & 590 & 550 & 550 \\
time advances & 704 & 682 & 682 & 643 & 610 & 610 \\
witness re-nomination attempts & 693 & 727 & 727 & 527 & 513 & 513 \\
suspension or resume step draws & 270 & 298 & 298 & 89 & 103 & 103 \\
no-op steps & 392 & 487 & 487 & 428 & 480 & 480 \\

%% file: supplement/static.tex
\begin{table}[htbp]
\centering
\caption{Slither 0.11.6 findings, de-duplicated across compilation units, before and after the fixes.}
\label{tab:static}
\footnotesize
\begin{tabularx}{\textwidth}{@{}p{3.0cm}p{1.1cm}ccL@{}}
\toprule
Detector & Impact & Initial & Current & Triage \\
\midrule
\input{tables/static.tex} 
\bottomrule
\end{tabularx}
\end{table}

%% file: tables/static.tex
\code{uninitialized-local} & Medium & 5 & 0 & Accumulators relied on implicit zero initialisation; now initialised explicitly. \\
\code{calls-loop} & Low & 6 & 8 & Registry reads inside loops over allocations, holders and (B2F) candidates; the registry is a trusted contract and loops are bounded by tranche size or the 32-candidate cap. \\
\code{missing-zero-check} & Low & 2 & 0 & \code{bindToken} accepted the zero address; it now reverts. \\
\code{timestamp} & Low & 15 & 15 & Validity windows, timelock and maturity compare \code{block.timestamp}; retained by design, since validator influence on timestamps is small relative to windows measured in days. \\

%% file: supplement/mutations.tex
\begin{table}[htbp]
\centering
\caption{Mutation checks. Each mutant weakens one boundary; ``killed'' means at least one test failed. TS: TypeScript functional suites before (initial) and after tests T19 and T20 were added.}
\label{tab:mutations}
\footnotesize
\begin{tabularx}{\textwidth}{@{}p{0.55cm}p{3.6cm}Lccc@{}}
\toprule
ID & File & Mutation & Fuzz & TS initial & TS final \\
\midrule
\input{tables/mutations.tex} 
\bottomrule
\end{tabularx}
\end{table}

%% file: tables/mutations.tex
M1 & \code{SukukPolicyEngine.sol} & issuance threshold lowered to 3299 bps & survived & killed & killed \\
M2 & \code{SukukPolicyEngine.sol} & allocation may exceed attested value by one unit & survived & survived & killed \\
M3 & \code{SukukPolicyEngine.sol} & allocation total may exceed proceeds by half & survived & survived & killed \\
M4 & \code{SukukToken.sol} & settlement may exceed a claim by one unit & killed & survived & survived \\
M5 & \code{EvidenceRegistry.sol} & claim windows up to 31 days accepted & killed & killed & killed \\
M6 & \code{EvidenceRegistry.sol} & equal sequence number accepted (replay) & killed & killed & killed \\

%% file: supplement/linecoverage.tex
\begin{table}[!htbp]
\centering
\caption{Line coverage of the production contracts in the instrumented run.}
\label{tab:linecoverage}
\footnotesize
\begin{tabular}{@{}lrrr@{}}
\toprule
Contract & Lines & Lines executed & Coverage (\%) \\
\midrule
\input{tables/coverage.tex} 
\bottomrule
\end{tabular}
\end{table}

%% file: tables/coverage.tex
\code{baseline/BasicCompliance.sol} & 7 & 6 & 85.7 \\
\code{policy/SukukPolicyEngine.sol} & 176 & 174 & 98.9 \\
\code{policy/SukukPolicyEngineFallback.sol} & 23 & 23 & 100.0 \\
\code{registry/EvidenceRegistry.sol} & 190 & 190 & 100.0 \\
\code{registry/IdentityRegistry.sol} & 9 & 8 & 88.9 \\
\code{token/SukukToken.sol} & 70 & 69 & 98.6 \\

%% file: supplement/costs.tex
\begin{table}[htbp]
\centering
\caption{Gas and hypothetical fees in native-token units. Fee $=$ gas $\times$ gas price $\times 10^{-9}$. The gas prices are assumptions, not observed market prices.}
\label{tab:costs}
\footnotesize
\begin{tabularx}{\textwidth}{@{}Lrrrr@{}}
\toprule
Item & Gas & \PriceLow{} gwei & \PriceMid{} gwei & \PriceHigh{} gwei \\
\midrule
\input{tables/costs.tex} 
\bottomrule
\end{tabularx}
\end{table}

%% file: tables/costs.tex
Added gas per ordinary transfer, B2 over B0 & 35{,}834 & 0.000036 & 0.000358 & 0.001792 \\
Added gas per ordinary transfer, B2F over B0 (witness valid) & 35{,}947 & 0.000036 & 0.000359 & 0.001797 \\
Added issuance gas per allocated asset, B2 (fitted slope) & 48{,}554 & 0.000049 & 0.000486 & 0.002428 \\
Renewal of one asset's three claims & 183{,}729 & 0.000184 & 0.001837 & 0.009186 \\
Renewal of the four claims the transfer check reads, per 30-day cycle & 241{,}852 & 0.000242 & 0.002419 & 0.012093 \\
Minimum liveness renewal per year (13 cycles) & 3{,}144{,}071 & 0.003144 & 0.031441 & 0.157204 \\
Full renewal per year, $n=3$ (13$\times$10 claims) & 7{,}921{,}035 & 0.007921 & 0.079210 & 0.396052 \\
Full renewal per year, $n=32$ (13$\times$97 claims) & 77{,}187{,}019 & 0.077187 & 0.771870 & 3.859351 \\

%% file: supplement/contracts.tex
\begin{table}[htbp]
\centering
\caption{Contracts of the prototype and the configurations that use them.}
\label{tab:contracts}
\footnotesize
\begin{tabularx}{\textwidth}{@{}P{4.3cm}LP{2.0cm}@{}}
\toprule
Contract & Responsibility & Used in \\
\midrule
\code{SukukToken} & ERC-20 with agent controls; one pre-check for every balance change; issuance against authorised tranches; redemption accounting & all \\
\code{IdentityRegistry} & investor eligibility flag, country code and expiry; a simplified stand-in for ERC-3643's identity registry & all \\
\code{BasicCompliance} & compliance contract that allows every issuance and transfer & B0 \\
\code{EvidenceRegistry} & attestor registry, EIP-712 verification, claim storage and eligibility views & B1, B2, B2F \\
\code{SukukPolicyEngine} & issuance and transfer policy, versioned governance, suspension, witness & B1, B2 \\
\code{SukukPolicyEngineFallback} & \code{SukukPolicyEngine} plus a candidate list and a bounded fallback scan & B2F \\
\code{InlineEvidenceHarness} & per-transfer verification of four signed claims, for measurement only & benchmark \\
\code{PropertyFuzz.t.sol} & Solidity fuzz properties for issuance, redemption accounting and claim windows & tests \\
\bottomrule
\end{tabularx}
\end{table}

%% file: supplement/bytecode.tex
\begin{table}[!htbp]
\centering
\caption{Runtime and initcode size of each contract, and runtime size as a share of the 24,576-byte limit of EIP-170.}
\label{tab:bytecode}
\footnotesize
\begin{tabular}{@{}llrrr@{}}
\toprule
Contract & Used in & Runtime bytes & Initcode bytes & Share of limit \\
\midrule
\input{tables/bytecode.tex} 
\bottomrule
\end{tabular}
\end{table}

%% file: tables/bytecode.tex
\code{SukukToken} & all & 7{,}271 & 8{,}406 & 29.6\% \\
\code{IdentityRegistry} & all & 1{,}929 & 2{,}262 & 7.8\% \\
\code{BasicCompliance} & B0 & 755 & 804 & 3.1\% \\
\code{EvidenceRegistry} & B1, B2, B2F & 12{,}300 & 13{,}636 & 50.0\% \\
\code{SukukPolicyEngine} & B1, B2 & 10{,}351 & 12{,}006 & 42.1\% \\
\code{SukukPolicyEngineFallback} & B2F & 10{,}741 & 12{,}406 & 43.7\% \\
\code{InlineEvidenceHarness} & benchmark only & 13{,}677 & 15{,}026 & 55.7\% \\

%% file: supplement/latency.tex
\begin{table}[!htbp]
\centering
\caption{Wall-clock latency on Hardhat's in-process simulated network (\LatencyTrials{} trials per operation and configuration). These timings describe the simulator on one machine and exclude block production and propagation on a public network.}
\label{tab:latency}
\footnotesize
\begin{tabular}{@{}llrrrr@{}}
\toprule
Operation & Config. & Median (ms) & IQR (ms) & Min (ms) & Max (ms) \\
\midrule
\input{tables/latency.tex} 
\bottomrule
\end{tabular}
\end{table}

%% file: tables/latency.tex
Transfer (send to receipt) & B0 & 2.54 & 2.38--2.84 & 2.08 & 25.28 \\
 & B1 & 3.08 & 2.85--3.29 & 2.46 & 20.79 \\
 & B2 & 3.38 & 3.08--3.77 & 2.66 & 22.42 \\
 & B2F & 3.02 & 2.75--3.56 & 2.48 & 18.02 \\
Pre-check (\code{eth\_call}) & B0 & 0.72 & 0.61--0.81 & 0.41 & 16.88 \\
 & B1 & 0.74 & 0.66--0.90 & 0.59 & 28.40 \\
 & B2 & 0.82 & 0.72--0.99 & 0.55 & 4.35 \\
 & B2F & 0.79 & 0.73--0.86 & 0.67 & 1.69 \\
Credential update (sign, submit, receipt) & B1 & 4.51 & 4.11--5.18 & 3.44 & 27.82 \\
 & B2 & 4.86 & 4.28--5.36 & 3.38 & 27.92 \\
 & B2F & 4.45 & 4.09--5.05 & 3.44 & 27.90 \\

%% file: supplement/types.tex
\begin{lstlisting}[language={},caption={EIP-712 type strings.},label={lst:types}]
Attestation(bytes32 instrumentId,uint8 claimType,bytes32 subjectId,uint8 code,
  bool flag,uint256 amount,uint32 policyVersion,bytes32 evidenceHash,
  uint64 seq,uint64 issuedAt,uint64 validUntil)

TrancheAllocation(bytes32 instrumentId,bytes32 trancheId,uint256 proceeds,
  bytes32[] assetIds,uint256[] amounts,uint32 policyVersion,
  bytes32 evidenceHash,uint64 issuedAt,uint64 validUntil)
\end{lstlisting}

%% file: supplement/claimtypes.tex
\begin{table}[htbp]
\centering
\caption{Claim types, required signer roles and fields used.}
\label{tab:claimtypes}
\footnotesize
\begin{tabularx}{\textwidth}{@{}clP{2.6cm}P{2.0cm}L@{}}
\toprule
Type & Name & Signer role & Subject & Fields used \\
\midrule
1 & Rights & legal verifier & asset & \code{code}: rights type (not interpreted on-chain); \code{flag}: encumbered \\
2 & Valuation & valuer & asset & \code{amount}: value in nominal units; zero means disposed or worthless \\
3 & Classification & Shariah adviser & asset & \code{code}: non-debt or debt; \code{policyVersion} \\
4 & Encumbrance & custodian & asset & \code{flag}: encumbered (negative evidence, does not lapse) \\
5 & Shariah status & Shariah adviser & instrument & \code{code}: endorsed or not endorsed; \code{policyVersion} \\
\bottomrule
\end{tabularx}
\end{table}

%% file: supplement/lifecycle.tex
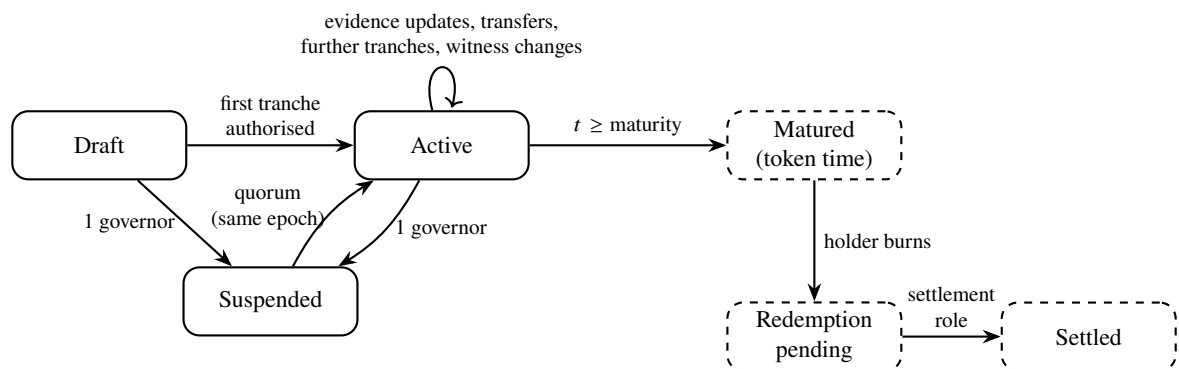
\begin{figure}[htbp]
\centering
\begin{tikzpicture}[
  font=\small, node distance=1.2cm and 2.2cm,
  st/.style={draw, thick, rounded corners=6pt, minimum width=2.3cm, minimum height=0.9cm, align=center},
  arr/.style={-{Stealth[length=2.4mm]}, thick},
  lab/.style={font=\footnotesize, align=center}
]
\node[st] (draft) {Draft};
\node[st, right=of draft] (active) {Active};
\node[st, below=1.6cm of $(draft)!0.5!(active)$] (susp) {Suspended};
\node[st, right=2.6cm of active, dashed] (mat) {Matured\\(token time)};
\node[st, below=1.6cm of mat, dashed] (pend) {Redemption\\pending};
\node[st, right=1.3cm of pend, dashed] (settled) {Settled};
\draw[arr] (draft) -- node[lab, above] {first tranche\\authorised} (active);
\draw[arr] (draft) -- node[lab, left] {1 governor} (susp);
\draw[arr] (active) to[bend left=15] node[lab, right] {1 governor} (susp);
\draw[arr] (susp) to[bend left=15] node[lab, left, pos=0.6] {quorum\\(same epoch)} (active);
\draw[arr] (active) -- node[lab, above] {$t \geq$ maturity} (mat);
\draw[arr] (mat) -- node[lab, right] {holder burns} (pend);
\draw[arr] (pend) -- node[lab, above] {settlement\\role} (settled);
\draw[arr] (active) edge[loop above] node[lab] {evidence updates, transfers,\\further tranches, witness changes} (active);
\end{tikzpicture}
\caption{Instrument lifecycle. Solid states are held by the policy engine; dashed states follow from the token's maturity time and redemption records. Resumption returns the engine to the state it held before suspension.}
\label{fig:lifecycle}
\end{figure}